\documentstyle[twocolumn,aas2pp4]{article}
\def\stacksymbols #1#2#3#4{\def\theguybelow{#2}
	\def\verticalposition{\lower#3pt}
	\def\spacingwithinsymbol{\baselineskip0pt\lineskip#4pt}
	\mathrel{\mathpalette\intermediary#1}}
\def\intermediary #1#2{\verticalposition\vbox{\spacingwithinsymbol
	\everycr={}\tabskip0pt
	\halign{$\mathsurround0pt#1\hfil##\hfil$\crcr#2\crcr
		\theguybelow\crcr}}}
\def\lta{\stacksymbols{<}{\sim}{2.5}{.2}}
\def\gta{\stacksymbols{>}{\sim}{3}{.5}}

\begin{document}
\title{EVOLUTION OF HOT GAS AND DARK HALOS IN GROUP-DOMINANT 
ELLIPTICAL GALAXIES: INFLUENCE OF COSMIC INFLOW$^1$}

\author{Fabrizio Brighenti$^{2,3}$ and William G. Mathews$^2$}

\affil{$^2$University of California Observatories/Lick Observatory,
Board of Studies in Astronomy and Astrophysics,
University of California, Santa Cruz, CA 95064\\
mathews@lick.ucsc.edu}

\affil{$^3$Dipartimento di Astronomia,
Universit\`a di Bologna,
via Zamboni 33,
Bologna 40126, Italy\\
brighenti@astbo3.bo.astro.it}






\vskip .2in

\begin{abstract}

Hot interstellar gas in elliptical galaxies has two sources:
mass lost from evolving stars and a much older 
component that accompanied galaxy formation or arrived 
subsequently by secondary cosmic infall toward the 
galaxy group containing the elliptical.
We present here an approximate but comprehensive study of the 
dynamical evolution of the hot 
gas in massive elliptical galaxies born 
into a simple flat universe.
Baryonic and dark matter are both conserved.
We use NGC 4472 as a prototypical massive elliptical 
having a well-observed hot interstellar medium.
We allow for star formation in a simple single burst 
using a Salpeter IMF 
but treat the gas dynamics in detail.
The galaxy has a de Vaucouleurs stellar core and a 
Navarro-Frenk-White dark halo surrounded by inflowing 
cosmic matter.

Using rather standard assumptions 
and parameters, we are able to 
successfully reproduce the gas density 
and temperature distributions -- $n(r)$ and $T(r)$ -- 
in the hot interstellar gas determined by recent X-ray observations.
Our model is sensitive to the baryon fraction of the universe,
the Type II supernova energy released per unit stellar 
mass, and the time of galaxy formation.
But there is some degeneracy; 
as each of these parameters is varied, 
the effect on model fits to $n(r)$ and $T(r)$ is similar. 
Nevertheless, 
secondary inflow of cosmic gas is essential for successful fits 
to $n(r)$ and $T(r)$.

Some gas is expelled from the stellar galactic core 
at early times
when the Type II supernova energy was released.
As a result, the present day baryonic fraction 
has a deep minimum in the outer galactic halo.
Interstellar gas that cooled since the time of 
maximum star formation 
cannot have all collected at the galactic center 
but must be spatially dispersed; otherwise both gas 
temperatures and stellar dispersions 
in the galactic center would be larger than those observed.

Finally, when relatively gas-rich, X-ray luminous models are 
spatially truncated at early times, simulating tidal 
events that may have occurred during galaxy group dynamics, 
the current locus of truncated 
models lies just along the $L_x$, X-ray size
correlation among well-observed ellipticals.
This is another striking confirmation of our model of 
elliptical evolution.

\end{abstract}



\section{INTRODUCTION}








In this paper we present a new, more comprehensive 
model for the evolution of hot gas 
in massive ellipticals and dominant ellipticals in galaxy groups.
For the first time we combine gas produced by stellar mass loss
with additional gas that flows into the galaxy regarded 
as a perturbation in the Hubble flow.
The radial distribution of gas density and temperature in bright 
ellipticals determined by 
recent X-ray observations can be understood 
only by combining these two sources of gas, stellar and cosmic.

Traditionally, it has been assumed that most or all of the hot 
gas observed in ellipticals arises as a result of mass loss from the 
aging stellar population,
but recent observations with the ROSAT satellite indicate 
a more complicated origin for the gas.
Davis \& White (1996) showed that the temperature of the hot gas 
is generally $\sim 1.5$ times hotter than the equivalent temperature 
of the stars $T_*$.
This implies that the gas is in virial equilibrium in a deeper, 
and therefore larger, potential than the stars.
Even more relevant, X-ray images from ROSAT 
have revealed very extended regions of X-ray emission surrounding
many massive ellipticals, extending far beyond the optical image.
Mathews and Brighenti (1998) have shown that the emission from 
these extended regions can dominate the overall X-ray luminosity.
This realization resolves a long-standing puzzle for 
elliptical galaxies: the enormous variation in 
X-ray luminosity $L_x$ in ellipticals having similar optical luminosity.
However, the total mass of hot gas in the outer regions of 
many of these galaxies is too large to be understood in terms
of stellar mass loss alone.
In Brighenti \& Mathews (1998) we showed that agreement 
with the observed hot gas density and temperature profiles is greatly
improved by assuming that a large mass of ``circumgalactic'' gas
existed around the galaxy at early times.
Much of the gas that filled the galaxy at the time of galaxy 
formation is still present today.
This raises the possibility that the hot gas in elliptical galaxies 
can reveal, like growth rings in a tree, the sequence of events that 
occurred long ago during the epoch of galaxy formation.

In this paper we extend this idea to consider the continuous 
inflow of gas into the galactic halo over the Hubble time.
This type of flow is a natural consequence of the large perturbation
in the Hubble flow that formed the galaxy and the small group 
in which it may reside.
Our approach in constructing these models is to adopt the 
simplest or most plausible procedure.
We assume for example 
that the evolution of the dark matter can be described 
by the normal self-similar flow in a perturbed flat universe, 
but as the central stationary dark halo grows in size, 
it adapts to a shape dictated by more detailed N-body calculations.
The baryonic gaseous component flows in with the dark matter 
but eventually shocks and compresses to the virial temperature 
$\sim 10^7$ K of the dark halo.
From the very beginning, as X-ray energy is radiated away, 
cold gas condenses at the center of the flow.
When a sufficient mass of baryons has accumulated, we form
the stellar distribution of the elliptical in a single event.
In this way we circumvent the complex dynamical and 
merging processes that 
occurred during the formation of the big elliptical.
At the same instant when the galaxy forms the remaining gas near 
the galaxy is heated by Type II supernovae and the galactic 
stars continue to evolve as a single burst thereafter.
In spite of the simplicity of this model,
the subsequent evolution of the gas, both that expelled from the 
stars and gas continuously flowing in toward the perturbation, 
is treated in a self-consistent, reasonably realistic manner.
This is confirmed by the agreement of our results with 
current X-ray observations.
It is gratifying that we can match so well 
the observed mass and entropy distribution 
in large ellipticals with our models and still conform to 
global cosmological and astrophysical constraints 
set by the conservation of baryonic mass, 
dark mass and supernova energy.

In this study we are concerned with establishing 
the global properties of hot gas in ellipticals 
as a consequence of gas dynamical evolution from the earliest times.
Therefore, in order to avoid unnecessary detail, we shall not discuss 
here the detailed metal abundance in the gas 
nor shall we consider the important effects of galactic rotation.

Although our results are generally applicable to all massive 
ellipticals and the group environment they inhabit,
as a rigorous test of the general success of our models it is useful
to use parameters specific to a particular well-observed 
galaxy; for this purpose we have chosen NGC 4472, the central
dominant elliptical in a subcluster in the Virgo cluster.

\section{PROPERTIES OF NGC 4472, A LUMINOUS ELLIPTICAL}

The X-ray  
surface brightness variation in massive ellipticals 
can be inverted to determine the spatial variation
of electron density with physical radius.
The radial plasma temperature profile can be 
similarly determined from variations in the 
apparent X-ray spectrum viewed in projection.
Since the hot interstellar gas is in near hydrostatic 
equilibrium, the total gravitating 
mass $M(r)$ can be found from 
the temperature and density profiles.
At present the best galaxies for this purpose 
are those relatively nearby and luminous ellipticals that 
have been imaged by both {\it Einstein} HRI 
and {\it ROSAT} HRI \& 
PSPC: NGC 4472, 4649, and 4636 (Brighenti \& Mathews 1997b).
Of these we chose NGC 4472 for comparison 
with our gas dynamical models since the X-ray 
data for NGC 4472 exists over the largest range, 
$\sim 0.016$ to $\sim 16$ effective radii where 
$r_e = 8.57$ kpc at a distance of 17 Mpc.
Brighenti \& Mathews (1997b) showed that 
the hot gas in NGC 4472 
is at rest in the {\it stellar} potential 
in the range $0.1 - 1$ $r_e$ where the total mass (assuming 
hydrostatic equilibrium) is identical
to the stellar mass predicted from 
the dynamical mass to light ratio based on a  
two-integral stellar distribution 
function (van der Marel 1991).
This excellent agreement indicates that gas pressure alone 
is sufficient to support the hot gas; no additional
pressure (magnetic, cosmic rays, turbulence, etc.) is needed 
nor indicated in $0.1 - 1$ $r_e$. 
Unfortunately, beyond a radius of $\sim$3.5 $r_e$ 
the X-ray image of NGC 4472 
becomes asymmetric probably as a result of an environmental 
ram pressure interaction or an interaction with the dwarf 
irregular galaxy UGC 7636 (Irwin \& Sarazin 1996; 
Irwin, Frayer, \& Sarazin 1997).
However, the azimuthally 
averaged X-ray surface brightness used
by these authors to determine the average
density profile is globally representative 
of luminous ellipticals.
NGC 4472 is a prototypical bright elliptical
that exhibits all of the characteristics of 
these massive galaxies.

The upper panel of Figure 1 shows the gas density 
(Trinchieri, Fabbiano \& Canizares 1986; Irwin \& Sarazin 1996) 
and temperature (Irwin and Sarazin 1996)
profiles in NGC 4472 together with analytic fits as 
described by Brighenti \& Mathews (1997b).
The corresponding total mass $M(r)$ is shown in the lower 
panel of Figure 1 
together with best fitting stellar and dark halo 
mass distributions.
The stellar mass is described with a de Vaucouleurs profile
with effective radius $r_e = 8.57$ kpc.
The stellar luminosity is $L_B = 7.89 \times 10^{10}$ 
$L_{B \odot}$ corresponding to a total stellar mass 
$M_* = 7.26 \times 10^{11}$ $M_{\odot}$ using 
a mass to light ratio of $M_{*}/L_B = 9.20$ (van der Marel 1991).
The dark halo in NGC 4472 dominates the total mass 
for $r \gta r_e$.
As shown in Figure 1 a dark halo described by the 
Navarro, Frenk \& White (1996) density profile (NFW halo) 
using a total halo mass $M_h = 4 \times 10^{13}$ $M_{\odot}$
fits the dark matter distribution quite well at large radii.
[A reservation about the applicability of the NFW profile
to NGC 4472 expressed by Brighenti \& Mathews 
(1997b) is now seen to be in error.]
The density distribution in an NFW halo is described by
$$\rho_h = {200 \over 3}~
{c^3 \over \ln (1 + c) - c/(1+c)}~
{\rho_H \over (r/r_s) [1 + (r/r_s)]^2 }$$
where $\rho_H$ is the current mean density of the universe,
$r_s = r_{gn}/c$ and 
$r_{gn}$ is the current virial radius of the halo 
where the density is $\sim 200 \rho_H$.
The concentration parameter, 
$c = 10.72 (M_h/3.3 \times 10^{13} M_{\odot})^{-0.1233}$, 
is based on Figure 8 of NFW.
All galactic parameters are scaled to our adopted 
distance to NGC 4472, $D = 17$ Mpc.

The agreement of 
the mass distribution of NGC 4472 
with an NFW halo plus a de Vaucouleurs core 
is good but not perfect,
particularly at the transition region near $\sim 10$ kpc.
The shape of the inner part of galaxy-sized dark halos 
is controversial, e.g. Kravtsov et al. (1998) 
predict a flatter density profile within the NFW characteristic
radius $r_s$; such a halo would improve the fit to the 
total mass in Figure 1.
However, any mass profile determined by N-body 
calculations in the absence of 
baryons will be centrally steepened and redistributed 
somewhat by the dissipative inward migration  
of the baryon component.
Nevertheless, in the calculations described here we shall
adopt the combined de Vaucouleurs and NFW mass distribution for 
NGC 4472. 
The relatively small discrepancy apparent 
in Figure 1 has essentially 
no significant influence on our gas dynamical models.

We have taken published profiles of the electron density $n(r)$ 
directly from the literature although 
Trinchieri, Fabbiano \& Canizares (1986) 
and Irwin \& Sarazin (1996) have made different assumptions about 
$T(r)$ and the metallicity variation $z(r)$ in the gas.
Since the indicated gas density depends on $T(r)$ and 
$z(r)$ as they influence the cooling rate, 
$n(r) \propto \Lambda(T,z)^{-1/2}$, the ``observed'' 
values of the gas density in Figure 1 could change 
somewhat when more definitive information on $T(r)$ and $z(r)$ 
becomes available.

\section{EVOLUTION OF THE DARK HALO BY SECONDARY INFALL}

For simplicity in estimating the growth of the dark halo in 
NGC 4472, we assume a critical $\Omega = 1$ universe and 
a current age $t_n = 13$ Gyrs corresponding to $H = 50$ km s$^{-1}$ 
Mpc$^{-1}$. 
The self-similar 
flow of dark matter toward an overdensity perturbation in 
a critical universe 
(Bertschinger 1985) consists of two regions: a smooth outer 
flow in which the Hubble flow reverses at a turnaround radius 
$r_{ta}(t)$ and an inner flow consisting of an ensemble of 
caustics or density cusps produced by orbital wrapping of the
collisionless dark fluid as it oscillates about the 
center of the perturbation. 
As infalling dark matter accumulates in the dark halo,
its mass is added to the outer part, 
leaving the central mass distribution essentially undisturbed.
Although newly arrived shells of dark matter pass through 
the central regions of the previously existing dark halo, 
they move 
very rapidly there and do not contribute appreciably to the 
time-averaged density except near the outermost caustic.
Averaging over the caustic peaks, the total dark mass density in 
Bertschinger's 
similarity solution varies as $\rho_d \propto r^{-9/4}$, but 
this is much too centrally peaked when 
compared to typical N-body simulations
(Navarro, Frenk \& White 1996).

We resolve this difficulty by assuming that an NFW halo
forms as a stationary core at the center of an outer smooth 
flow described by Bertschinger's similarity solution.
The outer flow must be attached to the stationary NFW halo 
so as to conserve total dark mass at all times.
In terms of the similarity variable
$$\lambda = { r \over r_{ta}(t) }$$
the outer solution can be fit quite precisely with
$${\cal M}(\lambda) = \lambda^3 ~ { a + b (\lambda / \lambda_1)^2
\over 1 + (\lambda / \lambda_1)^2 }$$
where $a = 3.4970$, $b = 5.8981$ and $\lambda_1 = 1.1296$.
This analytic fit to the smooth 
outer flow onto a point mass 
is exact at $\lambda = 0$, 1 and $\infty$.

The current 
virial radius $r_{gn}$ within the self-similar dark halo 
corresponds to an overdensity of $\sim 200$, i.e.
$$r_{gn} = \left( {3 M_n \over 4 \pi 200 \rho_{Hn}}
\right)^{1/3}$$
where $\rho_{Hn} = 1/6 \pi G t_n^2$ is the current total 
cosmic density and $M_n = 4 \times 10^{13}$ $M_{\odot}$ 
is taken to be the current virial mass.
In the self-similar solution the approximately stationary 
halo confronts the secondary inflow at the 
radius of the outermost caustic $r_{cn} = 1.528 r_{gn}$ 
at time $t_n$.
The current turnaround radius $r_{tan} = r_{ta}(t_n)$ can be 
found by insisting that the outer self-similar solution 
meet the NFW halo at radius $r_{cn}$ where the interior 
masses of the two mass distributions must agree: 
$${4 \over 3} \pi r_{tan}^3 \rho_{Hn} {\cal M}(r_{cn}/r_{tan})
= M_n { f(r_{cn}/r_s) \over f(c) } $$
where 
$$f(x) = \ln (1 + x) - {x \over 1 + x}$$
describes the NFW mass profile.
With our chosen parameters, the current turnaround 
radius is $r_{tan} = 1.096 \times 10^{25}$ cm.
By similarity arguments, the turnaround radius 
at earlier times is
$r_{ta}(t) = r_{tan} (t/t_n)^{8/9}$.

Finally, the radius $r_c$ at which the NFW halo 
meets the secondary inflow is found by solving 
$${4 \over 3} \pi [r_{ta}(t)]^3 \rho_{H}(t) 
{\cal M}[r_{c}(t)/r_{ta}(t)]
= M_n { f[r_{c}(t)/r_s] \over f(c) } $$
for $r_c(t)$, keeping $r_s$ constant.
Although ${\cal M}(\lambda)$ is unvarying with time 
in the similarity solution, we have relaxed this condition 
in the expression above to achieve a fit to the NFW halo.
The total mass distribution in the stationary 
halo and the external flow
are illustrated at several times in Figure 2.

\section{BARYONS AND SINGLE BURST STAR FORMATION}

Within the context of the $\Omega = 1$ cosmology the 
baryonic contribution to the total density $\Omega_b$ 
is constrained by primordial nucleosynthesis to 
a narrow range,
$$0.04 \lta \Omega_b (H/50)^2 \lta  0.06$$
(Walker et al. 1991).
For a given value of $\Omega_b$ the mass distribution 
in Figure 2, multiplied by $(1 - \Omega_b)$, represents 
the varying potential of dark matter. 
Provided the temperature of the baryonic gas is 
much less than the virial temperature of the dark halo, 
$T \lta 10^6$ K, cosmic gas will flow inward toward the 
galaxy precisely following the dark matter until the gas
encounters a shock front.
Within this accretion shock the mass distribution of the 
baryonic component is determined by gas dynamics and 
the assumed model for star formation. 

Interior to the outward facing shock the cosmic gas is 
heated to $T \sim 10^7$ K, similar to the virial temperature 
of the dark halo. 
After a few Gyrs a large mass of gas will have collected 
and radiatively cooled within the dark matter halo; 
it is natural to assume that this gas forms the 
old stellar system observed today.
In our very simple model for star formation we assume that 
all the stars in NGC 4472 form as a single burst at 
time $t_*$.
One possible choice of $t_*$ would be to assume that stars form 
from the cooled gas as as soon as a mass of cold gas equals 
the current stellar mass of NGC 4472.
However, dynamic considerations suggest that 
the de Vaucouleurs stellar profile results 
from the merging of stellar systems that must have formed 
prior to the merging events, so stars are likely to 
have predated NGC 4472.
Furthermore, 
gas cools more efficiently and can form stars somewhat sooner in 
an earlier generation of small galaxies in which the dark matter 
and cooling gas are more concentrated than in NGC 4472.
For these reasons 
we regard the time of star formation $t_*$ 
as a variable parameter in our models; if the total 
mass of radiatively cooled gas is less than 
the mass $M_*$ of NGC 4472 at 
time $t_*$, some of the 
hot gas is used to make up the deficiency since it could 
have cooled by this time in smaller pre-galactic stellar systems.
At the time of star formation the gas 
density is reduced by a factor $[M_g(r_{sh}) - M_*]/M_g(r_{sh})$ 
where $M_{g}(r_{sh})$ is the total (all of the cold and perhaps 
some hot) 
gas mass within the shock radius $r_{sh}$ just before $t_*$.
Experience has shown that this artificial mass 
rearrangement at $t_*$ results in a 
relatively short-lived gas dynamical transient.

Soon after time $t_*$ massive stars ($m > 8$ $M_{\odot}$) 
begin to produce SNII explosions which heat
the remaining gas considerably.
We deposit this SNII energy instantaneously 
at time $t_*$ by increasing the specific thermal energy density 
$\varepsilon (r)$ by an appropriate amount everywhere 
within the shock radius at $t_*$.
The possibility
that some SNII events predated the physical formation of NGC 4472 
will not be considered here. 
The combined energy of the SNII explosions is sufficiently 
large to drive an outflow from the galaxy and the associated
galaxy group, creating a second 
outward moving shock that moves out beyond the accretion 
shock described above.

As soon as stars form they begin to lose mass 
by normal stellar evolution.
Most of this stellar mass loss occurs at early times shortly 
after $t_*$.
Unlike more realistic, rotating galaxies in which radiatively 
cooled gas forms into disks of size $r_{disk} \gta r_e$ 
(Brighenti \& Mathews 1996; 1997a), in the purely spherical models
discussed here cold gas can only form at the very center.
From $t_*$ until the present time enough cold gas can form at the 
center of the galaxy to influence 
the gravitational potential there, causing the gas temperature 
within $r_e$ to rise considerably above values observed. 
We avoid this possibility here by assuming that the total 
mass of gas that cools after time $t_*$
also forms into stars distributed at all radii
within the stellar galaxy. 
The mass of stars that is assumed to 
form in this diffuse manner after time $t_*$ is 
much less than the current stellar mass of NGC 4472 and can
be ignored in evaluating the global galactic potential. 
Therefore to implement this assumption, we simply ignore the self gravity
of the cooled gas at the galactic center.
We also ignore the relatively small SNII contribution 
resulting from this type of additional star formation.

Although our model for star formation is 
obviously {\it ad hoc} and oversimplified,
we believe that the subsequent dynamics of the hot gas, which 
is our main interest here, is modeled reasonably well.

\section{GAS DYNAMICS}

The standard gas dynamics equations that describe the evolution
of hot interstellar gas in ellipticals are the usual 
conservation equations with additional source and sink terms:
$${ \partial \rho \over \partial t}
+ {1 \over r^2} { \partial \over \partial r}
\left( r^2 \rho u \right) = \alpha \rho_*,$$
$$\rho \left( { \partial u \over \partial t}
+ u { \partial u \over \partial r} \right)
= - { \partial P \over \partial r}
- \rho {G M_{tot}(r) \over r^2} - \alpha \rho_* u,$$
and
$$ \rho {d \varepsilon \over dt } =
{P \over \rho} {d \rho \over d t}
- { \rho^2 \Lambda \over m_p^2}
+ \alpha \rho_*
\left[ \varepsilon_o - \varepsilon - {P \over \rho}
+ {u^2 \over 2} \right]$$
where $\varepsilon = 3 k T / 2 \mu m_p$ is the
specific thermal energy;
$\mu = 0.62$ is the molecular weight and $m_p$ the proton mass.
$M_{tot}(r)$ is the total mass of stars, dark matter 
and hot gas within radius $r$.

All three equations involve source terms describing
the rate $\alpha \rho_*$ (gm s$^{-1}$ cm$^{-3}$) that 
gas is ejected from evolving stars and supernovae,
i.e. $\alpha = \alpha_* + \alpha_{sn}$. 
To estimate $\alpha_*(t)$ we assume a single burst of 
star formation with a power law 
initial mass function (IMF) between lower and 
upper mass limits $m_{\ell}$ and $m_u$. 
Using the procedure described by Mathews (1989) 
we find 
$$\alpha_*(t) = \alpha_*(t_n - t_*) 
[t/(t_n - t_*)]^{-s(t)}$$
where $t_n = 13$ Gyrs is the current time.
For a Salpeter IMF of slope $x = 1.35$ 
with $m_{\ell} = 0.08$ and $m_u = 100$ $M_{\odot}$ we find
$\alpha_*(t_n - t_*) = 5.6 \times 10^{-20}$ s$^{-1}$ 
and $s(t)$ varies slowly with time 
($s =$ 0.83, 0.97, 1.11, 1.34 at 
0.001, 0.1, 1, 11 Gyrs respectively). 

In the thermal energy equation $\Lambda(T,z)$ is the 
optically thin radiative cooling coefficient which
is a function of temperature and metallicity
(Sutherland \& Dopita 1993).
Thermal energy lost from the gas by radiation is 
largely restored by the gravitational compression of the gas 
as it flows deeper into the galaxy.
Radiative cooling eventually 
dominates in the densest parts of the flow.
In the computed models we allow the gas to cool 
to $10$ K.
In addition we assume that the interstellar gas
is heated by the dissipation of the orbital energy 
of mass-losing stars and by supernovae.
The mean gas injection energy is
$\varepsilon_o = 3 k T_o /2 \mu m_p$ where
$T_o = (\alpha_* T_* + \alpha_{sn} T_{sn})/\alpha$.
The stellar temperature $T_*(r)$ is found by 
solving the Jeans equation for a de Vaucouleurs 
profile with additional dark matter beyond $r_e$.
Type Ia supernova heating is assumed to be distributed 
smoothly in the gas; we ignore the buoyancy of 
gas locally heated by supernovae that may transport 
energy and metals to larger galactic radii.
The heating by supernovae is described by multiplying 
the specific mass loss rate of supernovae $\alpha_{sn}$ 
by the characteristic temperature of the ejected 
mass $m_{sn}$, $T_{sn} = m_p E_{sn} / 3 k m_{sn}$, i.e.
$$\alpha_{sn} T_{sn} =
2.13 \times 10^{-8}~ {\rm SNu}(t)~ (E_{sn}/10^{51} {\rm ergs})~$$
$$h^{-1.7}~ (L_B/L_{B \odot})^{-0.35} ~~~ {\rm K}~{\rm s}^{-1}$$
where $h \equiv H/100 = 0.5$ is the reduced Hubble constant and we
adopt $E_{sn} = 10^{51}$ ergs as the typical energy
released in both Type Ia and Type II supernovae.
The Type Ia supernova rate is represented in SNu units, 
the number of supernovae in 100 years expected from 
stars of total luminosity $10^{10} L_{B \odot}$.
The dependence $\alpha_{sn} \propto L_B^{-0.35}$ 
derives from the mass to light ratio for ellipticals 
given by van der Marel (1991),
$M_{*t}/L_B = 2.98 \times 10^{-3} L_B^{0.35} h^{1.7}$ 
in solar units.

The equations are solved in one dimensional
spherical symmetry using an appropriately 
modified version of the Eulerian code ZEUS.
Spatial zones increase logarithmically and extend far 
out beyond the turnaround radius where the cosmic 
gas is supersonically receding from the galaxy.
Therefore we can assume ``outflow'' boundary conditions 
at the outermost spherical zone, confident that no 
disturbance at the boundary can propagate back into the 
galactic flow.
For our purposes here it is not necessary to consider the 
detailed thermal physics of the intergalactic gas; 
we simply assume that the temperature of this gas 
is isothermal at $T = 10$ K until it passes 
through the accretion shock.
Although the intergalactic temperature is expected to 
be greater than this (due to photoelectric heating by 
quasar radiation for example), the value of the 
intergalactic gas temperature makes no difference 
in our calculations as long as it is small compared to 
the virial temperature of the dark halo of NGC 4472, 
i.e. $\lta 10^6$ K.

\section{SUPERNOVA HEATING}

Supernovae in ellipticals today are infrequent 
and are all of Type Ia.
Cappellaro et al. (1997) find a current rate of 
$0.066 \pm 0.027 (H/50)^2$ in SNu units.
The frequency of SNIa in the past is currently unknown 
although it seems reasonable to assume that it has 
been decreasing over cosmic time,
$${\rm SNu}(t) = {\rm SNu}(t_n) (t / t_n)^{-p}.$$
Ciotti et al. (1991) recognized that 
the relative rates of stellar mass ejection ($\sim t^{-s}$)
and Type Ia supernova ($\sim t^{-p}$)
determine the gas dynamical 
history of hot interstellar gas in ellipticals.
For example if 
$p > s$ then galactic winds may occur at early times
but if $p < s$ winds may occur at late times.
However, in the presence of large amounts of 
additional cosmic gas flowing into the galaxy, 
galactic winds driven by Type Ia supernovae 
will tend to be suppressed.
We assume here that $p = 1$ and SNu$(t_n)$ = 0.03
so Type Ia supernovae are unable to generate winds by themselves.

However, the enormous energy released by supernovae 
of Type II produced by massive stars at very early times 
can drive an outflow from the galaxy 
even in the presence of secondary infall.
With a single burst, power law IMF described by 
$$\phi(m) dm = \phi_o m^{-(1+x)} dm~~~~~~m_{\ell} < m < m_u$$
the number of SNII per unit stellar mass formed is
$$ \eta_{II} = { N_{II} \over M_{*t}} = {x - 1 \over x}
{ m_8^{-x} - m_u^{-x} \over m_{\ell}^{1-x} - m_u^{1-x}}$$
where $M_{*t}$ is the total initial mass of stars.
All stars greater than eight solar masses ($m > m_8$) are 
assumed to become SNII.
For the Salpeter IMF discussed above ($x = 1.35$, $m_{\ell} = 0.08$,
$m_u = 100$) we find $\eta_{II} = 6.81 \times 10^{-3}$ SNII 
per $M_{\odot}$.
The specific supernova number $\eta_{II}$ is
sensitive both to the IMF slope and the mass limits;
for example 
$\eta_{II} = 11.6 \times 10^{-3}$ for $x,m_{\ell},m_u = 1.35, 
0.3, 100$ and $\eta_{II} = 16.1 \times 10^{-3}$ for 
$x,m_{\ell},m_u = 1., 0.08, 100$. 
We conclude that $\eta_{II}$ is uncertain by a factor of 
at least 2 or 3.
If $\eta_{II} = 6.81 \times 10^{-3}$, the total 
amount of SNII energy produced during the single burst 
formation of NGC 4472 ($M_{*t} = 7.26 \times 10^{11}$ $M_{\odot}$)
is $4.9 \times 10^{60}$ ergs assuming $E_{sn} = 10^{51}$ ergs 
per supernova.
An additional uncertainty is the efficiency $\epsilon_{sn}$ of 
communicating this SNII energy to the hot interstellar gas; if 
massive stars form near cold, dusty gas clouds 
a considerably fraction of SNII 
energy could be radiated away by dust and dense gas.
In galaxy formation simulations it is assumed 
that $\epsilon_{sn} \sim 0.1 - 0.2$ 
of the supernova energy is returned to 
the interstellar gas (e.g. Kauffmann, White \& Guiderdoni 1993),
but in general $\epsilon_{sn}$ must be regarded as 
an adjustable parameter.

\section{THE STANDARD MODEL}

Our objective is to solve the gas dynamical equations from 
the earliest times to the present ($t_n = 13$ Gyrs) and identify 
model parameters that result in gas density and 
temperature profiles similar to those observed today.  
While the number of uncertain parameters is formidable -- 
$\Omega$, $\Omega_b$, $t_*$, $t_n$, IMF, $\Lambda(T,z)$, SNII energy, 
$\epsilon_{sn}$, etc. -- there 
are also robust constraints set by cosmological 
density variations and the global conservation of 
dark mass, baryonic mass and energy. 

We begin by describing the results of our
``standard'' model. 
The standard model is not chosen for its perfect agreement 
with $n(r)$ and $T(r)$ observed in NGC 4472, 
although that agreement is quite satisfactory, 
but because the parameters 
and procedures used in the standard model 
are the most plausible and least controversial.
In our standard model we assume $\Omega_b = 0.05$ and form 
the galaxy at time $t_* = 2$ Gyrs. 
A Salpeter IMF is used with $m_{\ell} = 0.08$ and 
$m_u = 100$ for which 
$\alpha_*(t_s) = 4.7 \times 10^{-20} 
(t/t_s)^{-1.26}$ s$^{-1}$
is an approximate, time-averaged value of 
the stellar mass loss rate and $t_s = t_n - t_*$ 
is the current age of the stars.
The IMF also determines the specific energy generated 
by SNII explosions,
$\eta_{std} = 6.81 \times 10^{-3}$ SNII per $M_{\odot}$, 
and we assume that this 
energy is fully converted to the hot gas, $\epsilon_{sn} = 1$.
Clearly the important parameter 
for any model is $\eta_{II} \epsilon_{sn}$. 
The current rate of Type Ia supernovae is $SNu(t_n) = 0.03$ 
and it has decreased with time as $t^{-p}$ with $p = 1$.
In the standard model the radiative cooling 
was varied with radius 
to simulate an increasing gas metallicity closer to the 
center of the galaxy, $\Lambda(T,r) = \Lambda(T)f(r)$
where $f(r) = \max[0.3,~2 - 1.7(r/r_z)]$ with $r_z = 40$ kpc.
Such a variation is consistent with the 
expected radial variation 
of gaseous metallicity and the metal-dependent cooling 
coefficients described by Sutherland \& Dopita (1993).
The function $f(r)$ is designed to match a  
radial abundance variation similar to that 
observed in the hot interstellar
gas in NGC 4472 (Forman et al. 1993; Matsushita 1997).

In Figure 3 we compare the computed gas density 
$n(r,t_n)$ and temperature $T(r,t_n)$ for 
the standard model with observations of NGC 4472.
The fit to the observed temperature is excellent and the 
the density agreement is acceptable, within a factor of 2 
of the observed values over a range of 100 in radius.
Recall that 
the ``observed'' density variation is also somewhat uncertain as 
discussed earlier.
Nevertheless, it appears that the gas density observed beyond 
10 kpc is higher than that of the standard model while 
the reverse is true within 10 kpc.
We have found that the first of these problems can be 
corrected while the second is more difficult.

Figure 4 shows in more detail the spatial variation of gas 
density, temperature, pressure and velocity over the full 
range of our calculations at three times including $t_n$.
It is remarkable how little the central flow profiles have 
changed since $t = 4$ Gyrs.
The fraction of the gas density within 100 kpc due to 
stellar mass loss decreases slowly with 
$\alpha_*(t)$. 
While new gas continues to flow into the galaxy at later times,
this gas accumulates largely in the outer regions which grow 
in size while the solution within $\sim100$ kpc 
remains in a quasi-steady state.
Notice that there are two shocks.
At time $t_n$ the main accretion 
shock is at $\log r_{kpc} = 3$ and it is preceded by 
a second, smaller shock 
at $\log r_{kpc} = 3.4$.
Beyond this outer shock,
at the turnaround radius, $\log r_{tan} = 3.6$, the perturbed 
Hubble flow velocity vanishes.
The influence of the outer shock can be clearly seen 
in all flow variables.
This outer shock begins at time $t_*$ when the SNII energy
is released; as it moves outward it creates a transient 
baryonic outflow that pushes some gas far out into 
the galactic halo and beyond. 
At later times radiative losses just behind the outer shock 
are significant.
The small increase in gas density before entering this 
outer shock is an expected feature of the perturbed 
Hubble flow.
This calculation was performed with an intergalactic 
gas temperature of $10$ K, but the flow pattern 
within the outer shock was unchanged when the 
calculation was repeated with 
intergalactic temperature increased to $10^4$ K.

The radial variation of the local baryonic fraction 
at time $t_n$ is shown in Figure 5 for 
the standard solution; the hot gaseous 
contribution to the baryon fraction is shown by the dashed line.
Interior to $\sim 100$ kpc most of the baryons are stellar, 
but the gaseous contribution becomes significant at 1 Mpc
and dominant beyond 10 Mpc.
The most interesting feature in this plot is the 
large baryonic minimum centered at $\sim 500$ kpc -- this 
feature is a relic of the gaseous evacuation created by 
the release of SNII energy.
Baryons formerly in this region were pushed outward, 
forming a region of baryonic
overdensity ($\Omega_b > 0.05$) 
just beyond a radius of 2.5 Mpc.
Clearly such a large radius is irrelevant for NGC 4472 since 
it would enclose the entire Virgo cluster.
However, the baryonic depletion is real 
and the rapid spatial variation of $\Omega_b/\Omega$ in the 
outer galactic halo 
must be considered in interpreting observed values
of the baryon fraction.
For example, {\it ROSAT} observations in NGC 4472 extend only to 
radius $\sim 150$ kpc where the local baryon 
fraction is seen to be very steeply declining in Figure 5.
Clearly, measurements of the baryon fraction at such galactic 
radii could easily be greater {\it or less} than the 
true cosmic value ($\Omega_b/\Omega = 0.05$ for our standard model).

We show in Figure 6 the time variation of the mass of hot gas, 
cooled gas and stars in the standard model.
At time $t_* = 2$ Gyrs all of the cold gas and some of the hot 
gas is consumed to form the single burst of 
stars in NGC 4472.
Thereafter the stellar mass is assumed to be constant, 
$M_*(t_*) = M_*(t_n)$;
we have ignored the small ($\sim 20$ percent) decrease in 
$M_*$ expected from stellar mass loss since time $t_n$.
Although hot gas ($T > 10^6$ K) continues to increase within 
the main accretion shock, the amount of hot gas within 
150 kpc is constant for $t \gta 3$ Gyrs, 
suggesting a quasi-steady state in this part of the flow.
Cold gas continues to form after time $t_*$ reaching a final mass 
$M_{cg}(t_n) = 4.8 \times 10^{10}$ $M_{\odot}$ at the 
end of the calculation.
While $M_{cg}(t_n)$ is
comparable to the total mass of hot gas within 150 kpc, 
it is only $\sim 7$ percent of the current stellar mass.
Unlike our earlier models in which we assumed that all 
of the hot interstellar gas is generated by stellar mass loss, 
the additional supply of cosmic gas at later times 
in the standard solution 
also results in more cold gas overall.

In computing the gravitational potential 
we have ignored the mass of cold gas, assuming that some or all 
of it contributed to later generations of stars,
formed throughout a significant volume of the galaxy. 
However, it is interesting to relax this assumption and 
include 
the gravity of cold gas which, in our idealized non-rotating 
models, forms a concentrated 
point mass at the very center of the galaxy.
This variant of the standard model is shown in Figure 7.
Compared to Figure 3, 
the density profile in Figure 7 is slightly improved within 
a few kpc, but 
gas within about 5 kpc of the galactic core in Figure 7 becomes 
heated as it compresses in the point mass potential 
established by the cold gas. 
This type of central heating, with temperatures 
extending to $2 - 3 \times 10^7$ K, has not 
been observed. 
We have examined the convective stability of this core of 
hot gas and find that it is stable apart from a small 
rather insignificant region $\lta 500$ pc from the very center.
However, the massive, compact cloud of cold gas that 
creates the central potential would
also increase the central stellar velocity dispersion 
in a manner similar to that of a central massive black hole.
But the total mass of cooled gas in this model,
$M_{cg}(t_n) = 4.8 \times 10^{10}$ $M_{\odot}$, is about 
15 times more massive than the largest known central black hole 
in elliptical galaxies, $3 \times 10^9$ $M_{\odot}$ for M87
(Kormendy \& Richstone 1995).
{\it We conclude that all the interstellar 
gas that cools since $t_*$ cannot reside 
within 2 or 3 kpc of the galactic center since its gravity 
would unrealistically heat both the stars and hot gas.}
We therefore propose that this cooled 
gas goes into a stellar subsystem or disk of dimension 
$\gta 3$ kpc where its gravitational influence is greatly lessened.
This assumption is consistent with our neglect of 
the gravity of the cold gas and, 
since $M_{cg}(t_n) \ll M_*$, 
the additional contribution of the cold gas mass to stars 
is relatively insignificant.
The stellar mass within 1 kpc $M_*(1kpc) = 4.1 \times 10^{10}$
$M_{\odot}$ is comparable to $M_{cg}(t_n)$,
but the stellar mass within 3 kpc is considerably larger,
$13 \times 10^{10}$ $M_{\odot}$.
Only about half of the cold gas mass was available to 
form additional stars several Gyrs after $t_*$, more recently 
cooled gas must be accommodated in another way if there 
is to be no optical evidence for recent star formation.
Disks are attractive, but 
we recognize that the formation of disks in rotating ellipticals 
(Brighenti \& Mathews 1996; 1997a) is accompanied by 
centrally flattened X-ray images that have not yet been observed. 
Options for the final disposition of cold gas 
will be clarified in the future when we include galactic 
rotation in models such as those described here.

To illustrate the critical importance of additional 
circumgalactic gas 
in determining the global properties of hot gas in ellipticals,
we show in Figure 8 the density and temperature variations at 
time $t_n$ for the standard model when the dark halo 
within 150 kpc is non-varying 
and the only source of interstellar gas is stellar mass loss.
The dramatic inadequacy of this model is seen in the density 
plot where the mass of hot gas is far too low in the outer galaxy.
But the gas temperature is also too low.
The central gas temperature in this model is consistent with 
observed temperatures only at the very center where 
the observed temperature is nearly equal to 
the stellar virial temperature $\langle T_* \rangle 
\approx 10^7$ K.

It is important to emphasize that the low temperature in this 
model cannot be brought into agreement with the observations 
by increasing the current and past rate of 
Type Ia (or Type II) supernovae.
There are two reasons for this.
First, increasing the SNIa rate also 
increases the interstellar iron abundance far above observed 
values (e.g. Loewenstein \& Mathews 1991).
Second, Brighenti \& Mathews (1998) found that the gas 
temperature profile at $t_n$ is surprisingly insensitive to 
increases in the SNIa rate until that rate becomes large 
enough to drive a galactic wind. 
When a wind develops the gas temperature rises 
far too much and the gas density 
falls very far below observed values at every galactic radius.
We could find no value of SNu$(t)$ that corrected the 
obvious temperature discrepancies like that shown in Figure 8.
Davis and White (1996) found that gas temperatures 
in all bright ellipticals exceed the stellar virial 
temperature $T_*$ by factors of $\sim$1.5, as 
we find in Figure 3. 
We conclude that the long term accumulation of cosmic gas 
in the dark halo -- that naturally results in both densities and 
temperatures similar to those observed -- 
is essential to understand the nature of hot gas in ellipticals 
today.

\section{VARIATIONS ON THE STANDARD MODEL}

We now briefly discuss additional models in which 
some of the parameters are varied from those used in the 
standard model; unless otherwise indicated 
the parameters are the same as 
those used in the standard model.

\subsection{Variation of $\Omega_b$}

Within the restrictions imposed by primordial 
nucleosynthesis, $\Omega_b (H/50)^2 = 0.05 \pm 0.01$ is tightly 
constrained.
Nevertheless, we find that our models are sensitive 
to these small changes in $\Omega_b$.
For example 
Figure 9 shows the density and temperature profiles at 
$t_n = 13$ Gyrs for $\Omega_b = 0.04$ and 0.06.
The overall agreement with 
the observed density and temperature profiles 
is noticeably degraded compared to the $\Omega_b = 0.05$ 
standard model described earlier.
Evidently the amount of cosmically inflowing gas 
that enters the galaxy has a profound
influence on its X-ray appearance.

The inflow of gas at the outer 
edge of the galaxy, 
some of which may have previously flowed out of the galaxy 
during the release of SNII energy, 
also strongly influences the metallicity 
of hot gas deep inside the galaxy. 
Interstellar gas within the stellar galaxy is a blend of 
gas that has come from stellar mass loss and from cosmic
inflow; its net metallicity is also a blend.
To illustrate the contribution from these 
two sources of gas and its dependence on 
$\Omega_b$, we show in Figure 10 
the fraction of gas at time $t_n$ that has
come from stellar mass loss within the galaxy since 
time $t_*$ when the SNII energy was released.
In the $\Omega_b = 0.04$ solution almost all of the gas 
within the galaxy must have about the same average 
metallicity as the stars.
In fact some of the gas lost from the stars in this model 
since $t_*$ has 
moved out beyond the stellar system which extends 
only to 100 kpc.
By contrast, only 0.7 and 0.3 of the gas at $r_e = 8.57$ kpc 
has come from post-$t_*$ stellar mass loss 
for $\Omega_b = 0.05$ or 0.06 respectively.
The blending of two sources of interstellar gas, 
having different metallicities, 
complicates further the current controversy regarding 
the apparent discrepancy of the gaseous iron abundance with 
that in the stars, $z_{Fe,gas} < z_{Fe,*}$
(Arimoto et al. 1997; Renzini 1998).
Our $\Omega_b = 0.04$ model may be 
inconsistent with this inequality.

\subsection{Variation of SNII Energy and $t_*$}

The discrepancies seen in Figure 9 must not 
be taken too seriously 
since some of the misalignment with observations
can be removed by adjusting the total 
SNII energy ($\eta_{II}$) released 
and the epoch $t_*$ of galaxy formation.
For example, when $\Omega_b = 0.06$ the excess gas density 
in Figure 9 can be expelled from the galaxy by increasing 
the total SNII energy released at time $t_*$
by $2$.
Alternatively, if the SNII energy is left unchanged, 
the $\Omega_b = 0.06$ model can be improved by forming 
the stars at an earlier time, $t_* \sim 1.5$ Gyr, since 
there is just enough baryonic gas available at that early time 
to construct the stellar galaxy of mass $M_*$.
Conversely, the $\Omega_b = 0.04$ solution in Figure 9 
can be improved by lowering the SNII energy by $2$ 
and by assuming $t_* \approx 3$ Gyrs.
Several models adjusted in this manner are shown in 
Figure 11.
As $t_*$ is increased the mass of gas within 
the accretion shock is greater 
so the post-SNII temperature is lower;
this follows since we suppose that the SNII energy is 
deposited within the shock radius at $t_*$. 
In all our models 
we have assumed complete efficiency in converting 
SNII energy to the hot gas, $\epsilon_{sn} = 1$, 
so we have probably somewhat overestimated the 
SNII heating. 
Each model depends on the product $\epsilon_{sn} \eta_{II}$ 
so if $\epsilon_{sn} < 1$ then $\eta_{II}$ would need 
to be increased by $1/\epsilon_{sn}$ for identical 
results.

In principle the degeneracy in the choice of the parameters 
$\Omega_b$, $\eta_{II}$ and $t_*$ can be 
partially removed by considering in detail the enrichment 
by elements produced in SNII events.
We shall not attempt this here, 
but simply note that to compensate 
a relatively small, $\pm 0.01$ change in $\Omega_b$ 
requires a factor of $\sim 2$ change in the SNII energy  
or in the mass available to be heated. 
It is therefore unlikely that successful models are 
possible if $\Omega_b$ is 
much greater or less than the bounds set by nucleosynthesis.

Finally, we have computed models in which SNII heating
is ignored altogether.
In such models the gas density at $t_n$ greatly exceeds
(by factors of 2 - 5) 
the observed gas density at all galactic radii.
The amount of cold gas at time $t_n$ is also unacceptably large,
$M_{cg}(t_n) \gta 10^{11}$ $M_{\odot}$.
We conclude that SNII heating plays a critical role in 
establishing the amount of hot gas observed today in 
massive elliptical galaxies.

\subsection{Variation of Cooling or Heating}

In our standard model we allow for a modest negative gradient in 
the gas metallicity by increasing $\Lambda(T,z)$ toward the center 
of the galaxy. 
When compared to models with spatially constant $\Lambda$,
the $d \Lambda/ d r < 0$ assumption slightly improves 
the fit to the observed gas density for $r \lta 10$ kpc.
There is an inconsistency with this approach since 
the observed densities have been determined assuming 
uniform metallicity.
However, the sensitivity of our models to reasonable
variations of $\Lambda$ is not large.

We have also explored the hypothesis advanced by 
Tucker \& David (1997), Binney \& Tabor (1995),
and Ciotti \& Ostriker (1997)
that an additional heating process is present in the hot 
gas that reduces the total amount of cold gas that forms 
over a Hubble time.
However, we have not succeeded in finding a heating 
source term for the thermal energy equation 
that significantly improved our models.
If a small amount of heating due to 
some spatially distributed source
is present, it is either rather ineffective or, when increased,
generates outward propagating waves or shocks that result in 
peculiar density and temperature profiles 
and unsatisfactory X-ray surface brightness distributions.
The limiting version of such models
are those in which there is no radiative cooling at all,
as if heating and cooling were in perfect balance,
as some have suggested.
When the standard model is recalculated with 
$\Lambda = 0$, for example, we find that the gas density 
in $r \lta 40$ kpc rises as $n \propto r^{-2.5}$, 
completely diverging from the observations;
the gas temperature is too low by $\sim 0.5$ in this
same region.
In general therefore, we find that 
additional sources of distributed heating can 
introduce undesirable side effects that tend to degrade the 
agreement with observations.

\subsection{Two kinds of ``dropout''}

The deviation between our theoretical models and the 
observed gas density is almost always in the sense that 
the model profile is too centrally peaked.
This type of deviation has been noticed for many years 
and has led to the hypothesis of distributed mass ``dropout''
in which hot gas is assumed to disappear from the flow
according to some radial dropout function 
that can be varied to obtain agreement with observations 
(e.g. Fabian \& Nulsen 1977; Sarazin \& Ashe 1989).
Originally it was hoped that the mass dropout could 
be understood as some type of thermal instability resulting
from an inhomogeneous flow. 
Unfortunately, detailed hydrodynamical calculations have 
shown that thermal instabilities are disrupted by a variety of 
gas dynamical  
instabilities before an appreciable amount of cold gas forms;
see Brighenti \& Mathews (1998) for a more detailed discussion 
of dropout and additional references.

Nevertheless, in view of the popularity of mass dropout 
models we illustrate in Figure 12 the density and temperature 
structure at time $t_n$ in the standard model when an 
additional sink term $-q(\rho/t_{do})$ is included on the right
side of the continuity equation. 
The model in Figure 12 is based on $q = 1$ and the dropout 
time is given by the local radiative cooling time 
$t_{do} = 5 m_p k T /2 \rho \mu \Lambda$ (Sarazin \& Ashe 1989).
The dense, cooling regions that are dropping out contribute 
substantially to the total emission, reducing the 
apparent gas temperature 
$T_{obs}$ below that of the the undisturbed ambient gas $T$; 
both temperatures are shown in Figure 12. 

However, almost the same density profile can be found if 
the stellar mass ejection rate $\alpha_*$ is uniformly 
lowered.
For comparison, 
a model based on lowering $\alpha_*$ by 0.5 at all times is 
shown with dotted lines in Figure 12.

A decrease in $\alpha_*(t)$ can be understood from two 
astrophysical arguments.
First it is possible to reduce $\alpha_*$ by altering the 
(power law) IMF parameters $x$, $m_{\ell}$ and $m_u$.
Unfortunately the stellar mass to light ratio also changes.
For example, for values of $x$, $m_{\ell}$ and $m_u$ 
that reduce $\alpha_*(t_n)$ by 2, the mass to 
light ratio $M/L_V$ must increase by $\sim 2$ 
for any choice of the 
three adjustable IMF parameters. 
Such a large stellar $M/L_V$ is inconsistent 
with the value determined by van der Marel (1991) and with 
the excellent fit to the X-ray data for NGC 4472 using 
his $M/L_V$ (Brighenti \& Mathews 1997b).
Second, $\alpha_*$ can be ``virtually'' lowered if some of the
gas ejected from evolving stars (having a normal IMF) 
never enters the hot gas phase.
This idea was discussed by Mathews (1990) but discarded 
because of the many hydrodynamic instabilities that can 
disrupt stellar ejecta, increase its surface area and 
promote dissipative and 
conductive heating by the hot gas environment.
However, if these difficulties can be avoided, 
the results in Figure 12 suggests that observations of 
NGC 4472 may be consistent with this second type of ``dropout''.

\section{TIDAL TRUNCATION AND THE RANGE OF X-RAY PROPERTIES 
AMONG ELLIPTICALS}

For many years X-ray astronomers have noted that the 
X-ray luminosities of ellipticals $L_x$ span a huge range 
for galaxies of similar optical luminosity $L_B$
(Eskridge, Fabbiano \& Kim 1995). 
In spite of much effort, no 
intrinsic property of elliptical galaxies had been 
found that correlated with this scatter (e.g. White \& 
Sarazin 1991).
However, in a recent paper (Mathews \& Brighenti 1998) we 
showed that 
$L_x/L_B$ correlates significantly with the 
physical size of the X-ray source, 
$L_x/L_B \propto (r_{ex}/r_e)^{0.60 \pm 0.30}$,
where $r_{ex}$ is the radius that contains half of the 
X-ray luminosity in projection.
The discovery of this correlation was not accidental, but was 
inspired by the hypothesis that halo gas and dark matter 
could be tidally traded among massive 
early type galaxies in small groups. 
Such a tidal exchange of halo material would result 
in large, tidally dominant and centrally located 
ellipticals having huge, extensive 
halos and secondary donor (giant) 
ellipticals with more modest, 
tidally truncated dark matter and hot gas halos.
Because of the short dynamical time scales 
in galaxy groups,
many authors have noted that 
groups are ideal for both the production of ellipticals 
by mergers 
and for promoting tidal warfare among the component galaxies
(Merritt 1985; Bode et al. 1994; Garciagomez et al. 1996;
Athanassoula et al. 1997; Dubinski 1997). 
It was this hypothesis that led us to seek the correlation 
between $L_x/L_B$ and $r_{ex}/r_e$.

We can now test this hypothesis by truncating the dark matter 
and hot gas halos in the models described in this paper.
Although we cannot describe fully realistic tidal effects 
with our spherically symmetric models, 
tidal truncations can be approximately simulated 
simply by removing all dark matter and hot gas
beyond a truncation radius $r_{tr}$ at some time $t_{tr}$ in 
the past.
For this purpose we begin not with a model that describes 
NGC 4472 at time $t_n$, but with an elliptical with 
somewhat higher $L_x$ and containing more hot gas.
This can be accomplished simply by lowering the  
SNII energy below that of our standard model.
Immediately 
after the truncation, hot gas just within $r_{tr}$ flows 
outward and a rarefaction wave passes into the 
interstellar gas at the sound speed.
After a sound crossing time 
from $r_{tr}$ to the center of galaxy, $\sim 1$ Gyr,
a new cooling flow equilibrium is established.

In Figure 13 we show as dashed lines 
the final density and temperature profiles at time $t_n$ 
for the standard model but modified with $\eta_{II} = 0.3 \eta_{std}$.
As expected, 
this model is overdense relative to NGC 4472 at large radii.
Also shown with solid lines are the current profiles of
a model that was truncated 
at $t_{tr} = 9$ Gyrs at radius $r_{tr} = 400$ kpc 
but is otherwise identical to the first model.
The outer density profile of NGC 4472 is nicely 
fit by the truncated model and agreement with observed 
gas temperatures is satisfactory.
{\it Since the size of the observed X-ray image 
of NGC 4472, $r_{ex} = 2.87 r_e$,  
is average for X-ray bright 
ellipticals of similar $L_B$, it is very likely 
that NGC 4472 suffered such a truncation in its past.}
This truncation could have been accomplished by nearby 
group galaxies which may not be as massive as 4472.
We find that the final $n(r)$ and $T(r)$ 
profiles are rather insensitive 
to the exact time when the truncation occurred.

Finally, in Figure 14 we show the trajectory in the 
($L_x/L_B,~r_{ex}/r_e$)-plane caused by 
truncations of this $\eta_{II} = 0.3 \eta_{std}$ model at 
five different radii $r_{tr}$, all at time $t_{tr} = 9$ Gyrs.
Data for the observed galaxies is provided in Table 1 of 
Mathews \& Brighenti (1998).
The large $\times$ at the right in Figure 14 
shows the final location of the untruncated,
gas-rich galaxy shown in Figure 13.
The position of each small $\times$
on the left shows the final locus of the truncated models. 
Note that models truncated at $r_{tr} = 300$ and 
(particularly) 400 kpc 
lie rather close to 
the observed locus of NGC 4472, as one would expect from Figure 13.
{\it As models of 
gas-rich ellipticals are truncated at different radii, 
they move precisely along the correlation 
among observed ellipticals.}
This most encouraging result strongly supports our 
hypothesis that the spread in the
($L_x/L_B,~r_{ex}/r_e$)-plane is generated by 
tidal exchanges of halo material in small groups.


\section{FURTHER REMARKS}

We have not considered here the detailed 
metal enrichment of the hot 
gas expected from Type Ia and II supernovae; 
this will be the subject of a future study.
Clearly, accurate observations and theoretical 
predictions of the radial variation of the 
abundance of iron, silicon and other elements 
in the hot gas will 
provide important additional constraints on 
the formation and evolution of elliptical galaxies.
In the absence of this detailed information, 
it is interesting to 
compare the total amount of iron produced 
by SNII in NGC 4472 with the 
iron present in the stellar system today.
According to Trager (1997) the ratio of iron to $\alpha$ 
elements is less than solar in most bright ellipticals, 
suggesting that the mean stellar iron abundance 
in NGC 4472 is about $\langle z_{Fe,*} \rangle \sim 0.7$.
The total mass of stellar iron in NGC 4472 is then
$M_{Fe} \approx \langle z_{Fe,*} \rangle M_*/1.4 =
6 \times 10^8$ $M_{\odot}$ where 1.4 is the ratio of 
total to hydrogen mass.
If $\sim 0.14$ $M_{\odot}$ of iron is produced in each
SNII event (Gibson, Loewenstein, \& Mushotzky 1998),
then Type II supernovae should produce a total iron mass of
$\sim 7 \times 10^8$ $M_{\odot}$, comparable 
but somewhat larger than 
the mass of iron observed in stars today.
Evidently the iron-producing capability of the 
Salpeter IMF is sufficient to account for all the 
iron in stars. 
In making this estimate we have assumed that
the iron enrichment due to Type Ia supernovae goes 
primarily into the hot gas phase rather than into stars;
the total amount of iron in the hot gas within 
$150$ kpc is rather 
small, $\sim 3 \times 10^7$ $M_{\odot}$.

In general the density profiles in 
our most successful models resemble 
power laws throughout most of the observed galaxy 
while the observed $n(r)$ suggests a double power law 
with a flatter slope within $\sim 16$ kpc.
As a result our models predict a steeper 
gas gradient within the central $\sim 16$ kpc than is observed.
It is this small deviation that the ``dropout'' hypothesis
was designed to correct.
While future observations will clarify this curious systematic 
deviation, it is possible that new physical effects become
important in $r \lta 10$ kpc.
We have already noted that a small, but realistic galactic
rotation could flatten the azimuthally-averaged
gas density profile in this region.
Another possibility within $\sim$1 kpc is the 
growth of magnetic pressure toward the galactic center.
Brighenti \& Mathews (1997b) have shown that the total mass 
predicted from hydrostatic models is systematically lower 
than the known 
stellar mass within the central kiloparsec in each of the 
three bright ellipticals studied with {\it Einstein} HRI.
This deviation can be understood if the gas pressure is 
assisted by an additional component of non-thermal pressure.
The required magnetic field is large -- we estimated 
$B \sim 10^{-4}$ gauss for NGC 4636 -- but not unexpected.
Mathews \& Brighenti (1997) showed that fields of this 
magnitude are a natural outcome of the ejection of fields by 
mass-losing stars.
Fields ejected in stellar envelopes 
are amplified by turbulent dynamo 
action in the interstellar medium and by the natural 
intensification that accompanies the inward 
flow of interstellar 
gas in which the magnetic pressure increases more rapidly 
than the gas pressure.
The importance of magnetic or other non-thermal pressure components
in the interstellar cores of ellipticals 
will be clarified when AXAF observations become available.


\section{CONCLUSIONS}

We have described here the overall evolution of massive elliptical
galaxies with particular attention to the hot interstellar gas.
Although the stellar systems in giant ellipticals are thought to 
be very old, the mixing of the stellar orbits masks many details 
of the dynamical and enrichment processes that occurred 
during the time of galaxy formation.
The metallicity gradient in the stars provides some clue 
to these early events but its interpretation is far from obvious.

In contrast, the hot interstellar gas in ellipticals today does 
retain a large amount of information about star formation and 
metal enrichment events that occurred during the earliest phases
of galaxy evolution.
Interstellar gas in ellipticals can help us solve 
one of the thorniest problems in modern studies of galaxy formation: 
the importance of ``feedback'' SNII energy on the gas.
Among the major feedback uncertainties are: the total amount 
of supernova energy involved, its efficiency in heating the gas, and
the time that SNII enrichment occurred relative to the dynamical
formation of the galaxies.
Regarding this last point, 
it is currently unclear if some fraction of the metals 
in the hot gas were made and 
ejected from an earlier generation of low mass galaxies 
before the formation of galaxy groups,
or if all the metals were made during the 
formation of group galaxies that merged soon thereafter 
to create the ellipticals observed today.

We believe that studies of the creation and evolution of hot gas 
in large ellipticals provide a unique insight into the feedback
problem and may help us unravel some of these complex events that 
occurred so long ago.
Many lines of astronomical evidence indicate that the formation 
sites of most massive ellipticals are small groups of galaxies 
that are very old.
Dynamical studies of the evolution of galaxies in
elliptical-forming groups
indicate that
ellipticals formed early in the dynamical evolution of these 
groups when mergers and tidal interactions were most likely;
if so, both the elliptical galaxy and the hot interstellar gas
surrounding it are very old.
Remarkably, the hot gas we see today still 
contains a record of these past events.
Since SNII enrichment is accompanied by an increase in entropy 
due to heating of the gas, enrichment and gas dynamics are closely 
linked.

In this study we have described 
the first comprehensive theoretical description 
of the evolution of hot gas in massive ellipticals that 
is fully consistent with important cosmological constraints.
The goal of our calculations is not merely to explain the total 
mass and energy in the interstellar medium, but to account for 
the detailed radial dependence of these quantities throughout 
the observable range.
In spite of the simplicity of our galaxy-forming model 
and our choice of a currently unfashionable flat cosmology, 
we have been remarkably successful in explaining X-ray
observations of bright ellipticals 
in a cosmological context that conserves baryonic and dark mass, 
allows for all energy sources involved, and includes the detailed 
interaction of gas with mass-losing stars.

We now summarize some of the main conclusions of the calculations 
we have discussed here:

(1) Hot gas density and temperature profiles in massive ellipticals 
cannot be understood solely with gas ejected from the old 
stellar population that accounts for the optical light observed 
today.

(2) Low interstellar gas temperatures in models without cosmic
inflow cannot be heated to the observed values by 
current and past Type Ia supernovae;
if this is attempted, strong galactic winds develop that are
incompatible with observation. 

(3) Excellent agreement with observed hot gas mass and thermal 
energy distributions -- $n(r)$ and $T(r)$ -- can be achieved 
by the addition of cosmic gas from secondary infall.
This gas shocks and compresses to the virial temperature of 
the galactic halo, accounting for the hot gas temperatures 
observed in all massive ellipticals.
As this gas flows deeper into the galaxy, it mixes with somewhat
cooler gas expelled from the stars, resulting in 
positive gas temperature gradients ($dT/dr > 0$) 
in $r \lta 3 r_e$ as observed in 
most or all bright ellipticals (see Fig. 1 of 
Brighenti \& Mathews 1997b).

(4) Within the framework of a flat cosmology, we can account 
for observed gas density and temperature profiles 
only for a rather narrow range in baryon mass, $\Omega_b \approx 
0.05 \pm 0.01$, the same range that is consistent with 
primordial nucleosynthesis.

(5) The fraction of gas near the optical effective radius that 
derives from stellar mass loss is strongly and inversely 
dependent on $\Omega_b$ when other parameters are held constant.

(6) However, our successful fits to observed $n(r)$ and $T(r)$ 
are degenerate in some of the parameters. 
For example, increasing $\Omega_b$ is approximately 
equivalent to (i) decreasing the SNII energy ($\eta_{II}$) 
or its efficiency in heating 
the gas ($\epsilon_{sn}$) or 
(ii) increasing the time $t_*$ that the bulk of 
galactic stars formed so that more gas (within the accretion 
shock) needs to be heated by SNII.
Both (i) and (ii) reduce the energy per gram delivered to 
the gas by SNII explosions.
This degeneracy is due in part to the simple, single burst 
assumption we have made in forming stars 
and our assumption that SNII energy is deposited within the 
shock radius at time $t_*$.
More complex models that consider the details of metal enrichment 
and other options for the spatial distribution of SNII energy 
may remove some or all of this degeneracy.

(7) The current fraction of mass in baryons is expected 
to have a large minimum surrounding 
large ellipticals and their associated groups.
This minimum results from an outward displacement of baryons 
that occurred when the energy of Type II supernovae was released.
As a result, 
observational measurements of $\Omega_b/\Omega$ within large 
galactic radii can either overestimate or underestimate
the true cosmic value of this ratio.

(8) Since the epoch of star formation the amount of gas that 
has cooled is comparable to the total mass of hot gas observed 
today.
This gas cannot all cool within the central few kiloparsecs 
since its mass would unrealistically heat both the gas and 
stars there.
However, even a tiny galactic rotation is sufficient to avoid this 
central concentration of cooled gas.

(9) Finally, when a relatively gas-rich version of one of our 
models is tidally truncated at some time in the past, the 
X-ray luminosity $L_x$ and the effective radius for X-ray emission 
$r_{ex}$ both decrease. 
The vector due to truncation in the ($L_x/L_B,~r_{ex}/r_e$)-plane 
lies almost exactly along the correlation discovered 
by Mathews \& Brighenti (1998).
This give additional support to the notion, described by 
Mathews \& Brighenti (1998), that the enormous 
spread in $L_x$ for fixed 
galactic $L_B$ is due in some large measure to the tidal 
exchange of halo gas and dark matter among galaxies 
within groups of galaxies.

\acknowledgments

We thank our Santa Cruz colleagues for enlightening remarks and 
helpful criticism.
Thanks also to Caryl Gronwall for determining some stellar mass to 
light ratios for us.
Our work on the evolution of hot gas in ellipticals is supported by
NASA grant NAG 5-3060 for which we are very grateful. In addition
FB is supported 
in part by Grant ARS-96-70 from the Agenzia Spaziale Italiana.





\clearpage

\vskip.1in
\figcaption[aasrexfig1.ps]{
{\it Upper panel:} Open circles show 
ROSAT observations of plasma temperature (upper curve) and density 
(lower curve). Filled circles are derived from 
{\it Einstein} HRI observations.
{\it Lower panel:} Solid curve is the total gravitating mass based
on assuming hydrostatic equilibrium;
Dotted line is a de Vaucouleurs fit to the stellar core;
Long dashed line is an NFW dark halo that fits the outer galaxy.
Dot-dashed line is the sum of the de Vaucouleurs and NFW mass 
profiles.
\label{fig1}}

\vskip.1in
\figcaption[aasrexfig2.ps]{
NFW dark halo (solid line) and mass in the secondary inflow 
(dashed lines) shown at four times: 0.5, 3.0, 8.0 and 13.0 Gyrs 
from left to right.
At large radii the mass increases as $r^3$.
\label{fig2}}

\vskip.1in
\figcaption[aasrexfig3.ps]{
{\it Upper panel:} gas density profile of the standard model
compared to observations of NGC 4472;
{\it Lower panel:} gas temperature profile of the standard model
compared to observations of NGC 4472.
\label{fig3}}

\vskip.1in
\figcaption[aasrexfig4.ps]{
Flow variables in the standard model 
at three times: 4 Gyrs (dotted lines), 
9 Gyrs (dashed lines) and 13 Gyrs (solid lines).
{\it Upper left:} gas density,
{\it Upper right:} gas pressure,
{\it Lower left:} gas velocity, and 
{\it Lower right:} gas temperature.
\label{fig4}}

\vskip.1in
\figcaption[aasrexfig5.ps]{
{\it Solid line:} Variation of the baryon fraction at time $t_n = 13$ Gyrs 
in the standard model.
{\it Dashed line:} The baryonic contribution of the gas.
As the radius increases the fraction becomes $\Omega_b = 0.05$.
\label{fig5}}

\vskip.1in
\figcaption[aasrexfig6.ps]{
History of various mass components in the standard model. 
The stellar system is created at time $t_* = 2$ Gyrs.
{\it Solid line:} Variation of the mass of hot gas within the 
accretion shock radius,
{\it Dashed line:} Variation of the mass of hot gas within 150 
kpc,
{\it Dotted line:} Variation of cold gas ($T < 10^5$ K),
{\it Dot-dashed line:} Variation of the mass in stars. 
\label{fig6}}

\vskip.1in
\figcaption[aasrexfig7.ps]{
Density and temperature profiles for the standard model 
(as in Fig. 3) but including the gravity of the cold gas,
assumed to be a central point mass.
{\it Upper panel:} gas density profile 
compared to observations of NGC 4472;
{\it Lower panel:} gas temperature profile 
compared to observations of NGC 4472.
\label{fig7}}

\vskip.1in
\figcaption[aasrexfig8.ps]{
Density and temperature profiles for the standard model 
(as in Fig. 3) but with stellar mass loss as the only 
source of interstellar gas.
{\it Upper panel:} gas density profile
compared to observations of NGC 4472;
{\it Lower panel:} gas temperature profile
compared to observations of NGC 4472.
\label{fig8}}

\vskip.1in
\figcaption[aasrexfig9.ps]{
Density and temperature profiles for models
similar to he standard model in Figure 3 but with 
$\Omega_b = 0.06$ (solid line) and 
$\Omega_b = 0.04$ (dashed line).
\label{fig9}}

\vskip.1in
\figcaption[aasrexfig10.ps]{
Fraction of gas ejected from galactic stars since 
time $t_*$ for the standard model with 
$\Omega_b = 0.05$ (solid line),
and for similar models computed with 
$\Omega_b = 0.04$ (dashed line) and 
$\Omega_b = 0.06$ (dotted line).
\label{fig10}}

\vskip.1in
\figcaption[aasrexfig11.ps]{
Several models with various values of 
$\Omega_b$, $t_*$ and $\eta_{II}$.
{\it Solid line:} 
$\Omega_b = 0.05$, $t_* = 2$ Gyrs and $\eta_{II} = 0.6 \eta_{std}$.
{\it Dashed line:} 
$\Omega_b = 0.04$, $t_* = 3$ Gyrs and $\eta_{II} = 0.5 \eta_{std}$.
{\it Dotted line:} 
$\Omega_b = 0.06$, $t_* = 2$ Gyrs and $\eta_{II} = 2 \eta_{std}$.
\label{fig11}}

\vskip.1in
\figcaption[aasrexfig12.ps]{
Density (upper panel) and temperature (lower panel) distributions
of dropout and low $\alpha_*$ models compared with observations 
of NGC 4472.
{\it Dashed line:} Dropout in the standard model with $q = 1$,
{\it Solid line:} Dropout in standard model with reduced SNII 
energy, $\eta_{II} = 0.6 \eta_{std}$,
In the dropout models the upper (thin) lines show the 
temperature variation
of the background flow and the lower (thick) line show the 
apparent temperature $T_{obs}$ allowing for the cooling regions.
{\it Dotted line:} Standard model with $\alpha_*$ reduced by 
a factor 0.5 at all times and galactic radii.
\label{fig12}}

\vskip.1in
\figcaption[aasrexfig13.ps]{
{\it Dashed line:} Density and temperature profiles 
at $t_n$ for a model 
similar to he standard model in Figure 3 but with
$\eta_{II} = 0.3 \eta_{std}$.
{\it Solid line:} A repeat calculation of the same 
model at $t_n$ which was truncated at radius $r_{tr} = 400$
kpc at time $t_{tr} = 9$ Gyrs.
\label{fig13}}

\vskip.1in
\figcaption[aasrexfig14.ps]{
A plot of the
($L_x/L_B,~r_{ex}/r_e$)-plane showing the correlation 
among observed ellipticals.
The large $\times$  at the right is the locus at $t_n$ of the 
untruncated model shown in Figure 13.
The small $\times$s that extend away from the untruncated 
solution toward the lower left show the locus 
at $t_n$ of models truncated at time $t_{tr} = 9$ Gyrs 
at five decreasing truncation radii $r_{tr} = 500$, 400, 
300, 200, and 100 kpc.
As $r_{tr}$ decreases the locus of the final position 
in the ($L_x/L_B,~r_{ex}/r_e$)-plane moves from the 
untruncated solution almost 
exactly through the observed correlation.
\label{fig14}}

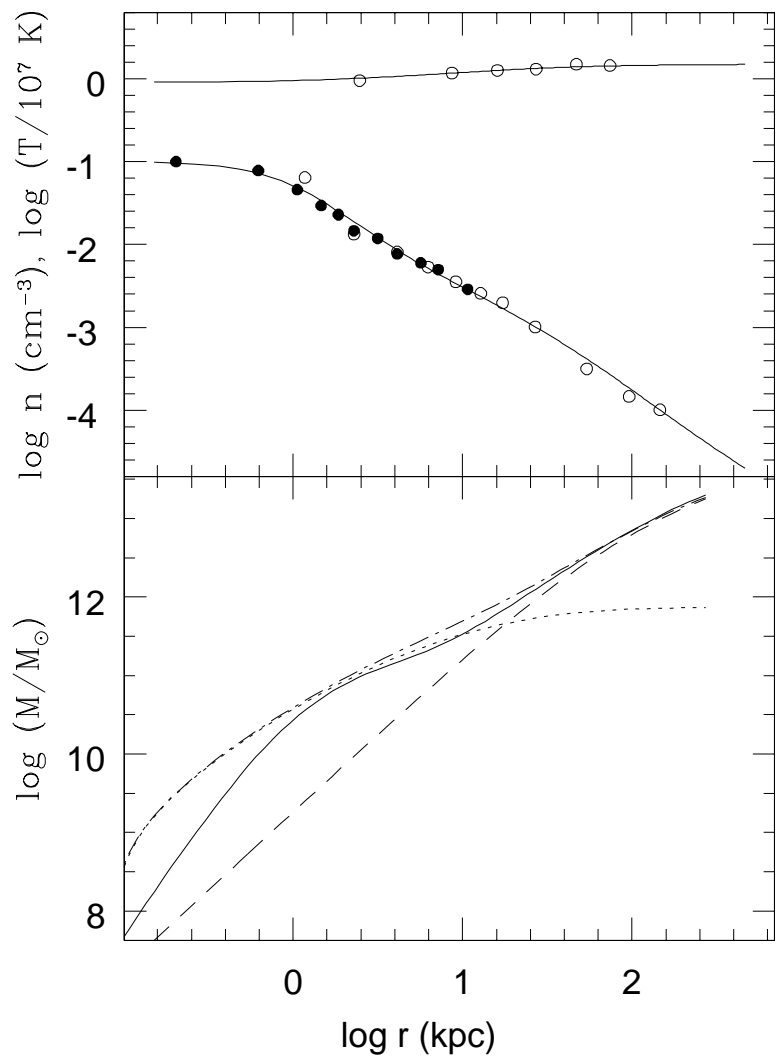
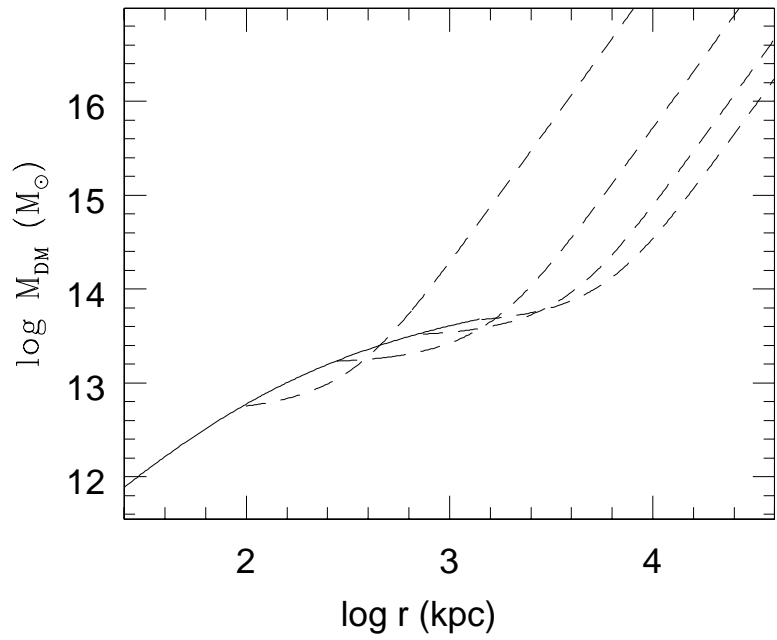
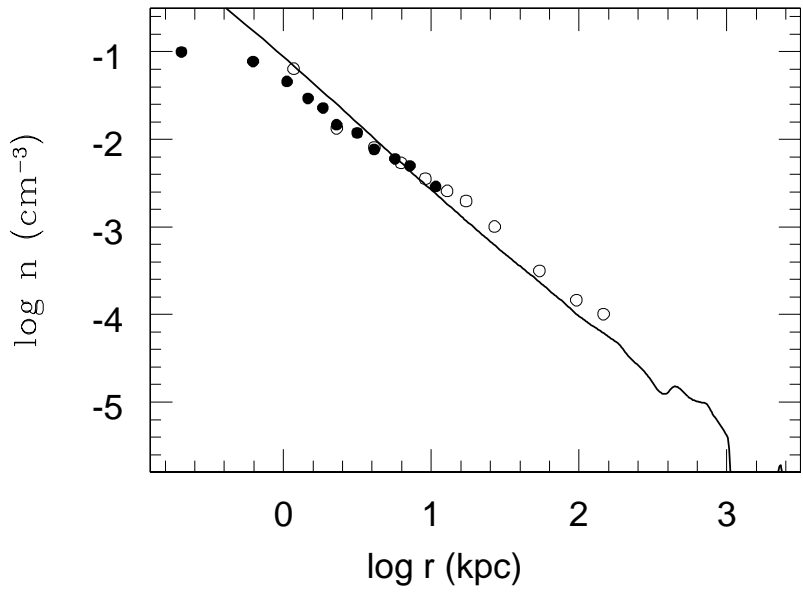
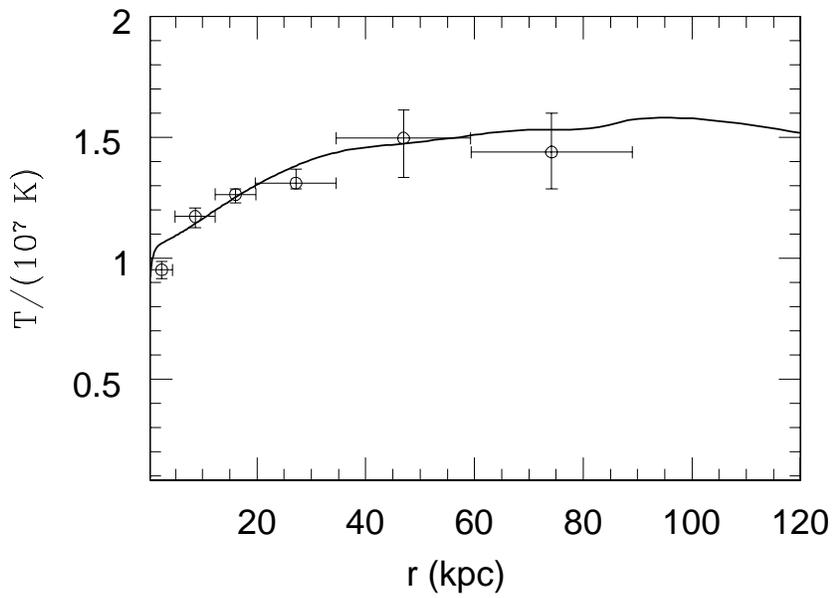
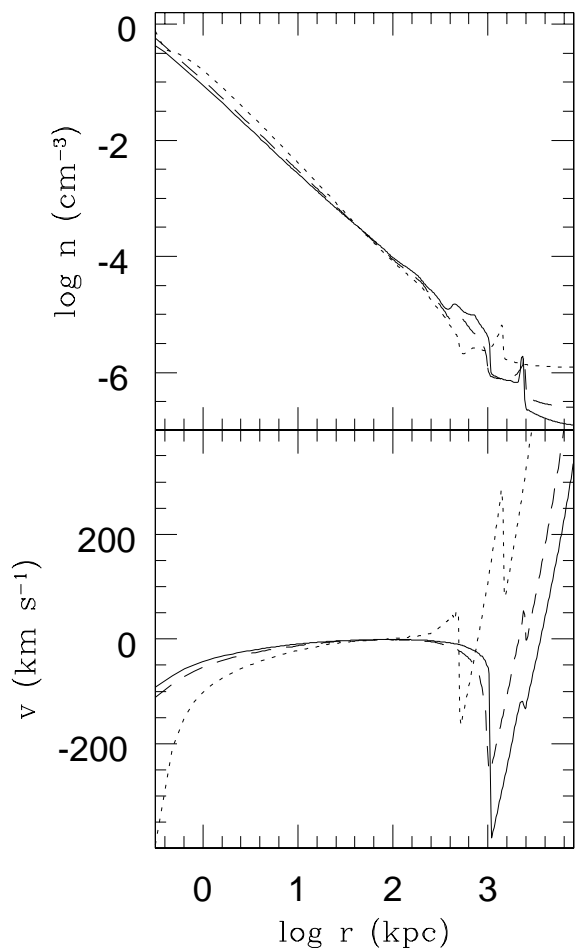
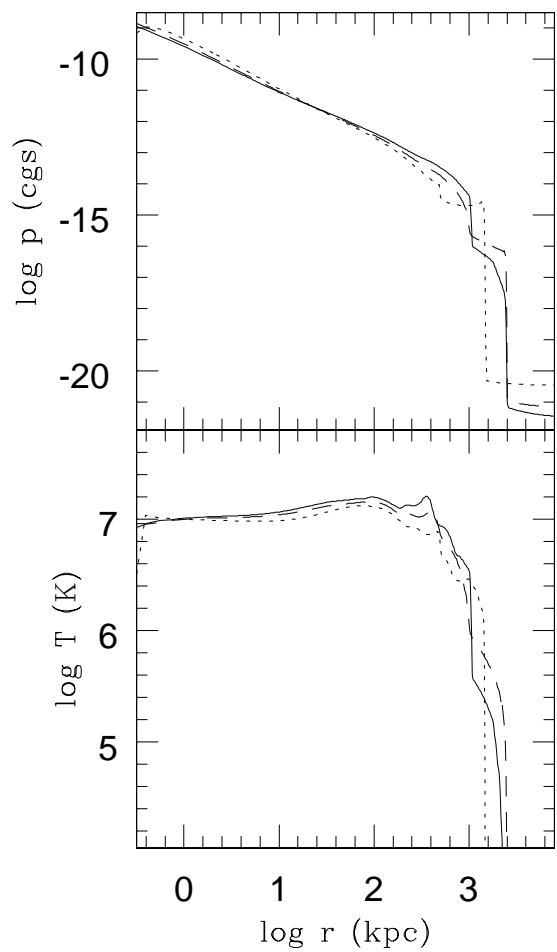
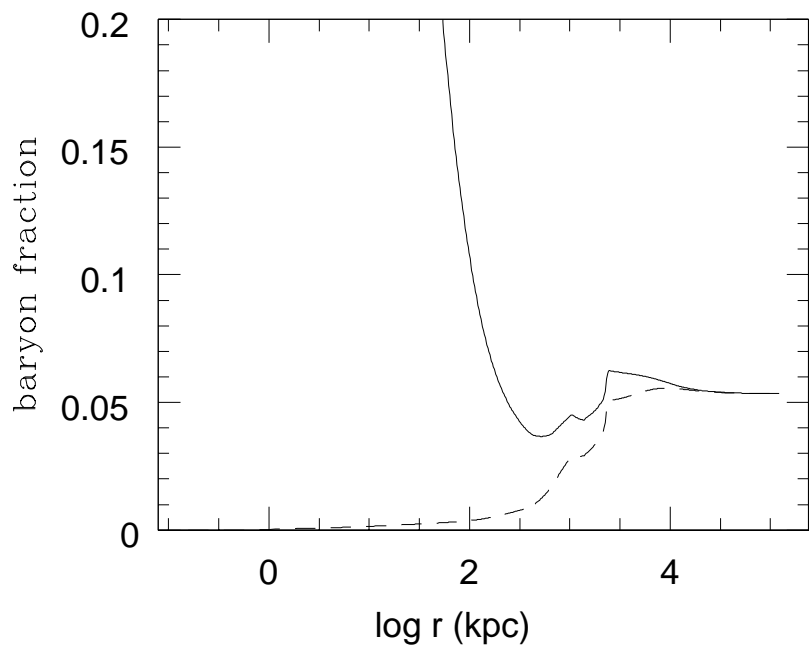
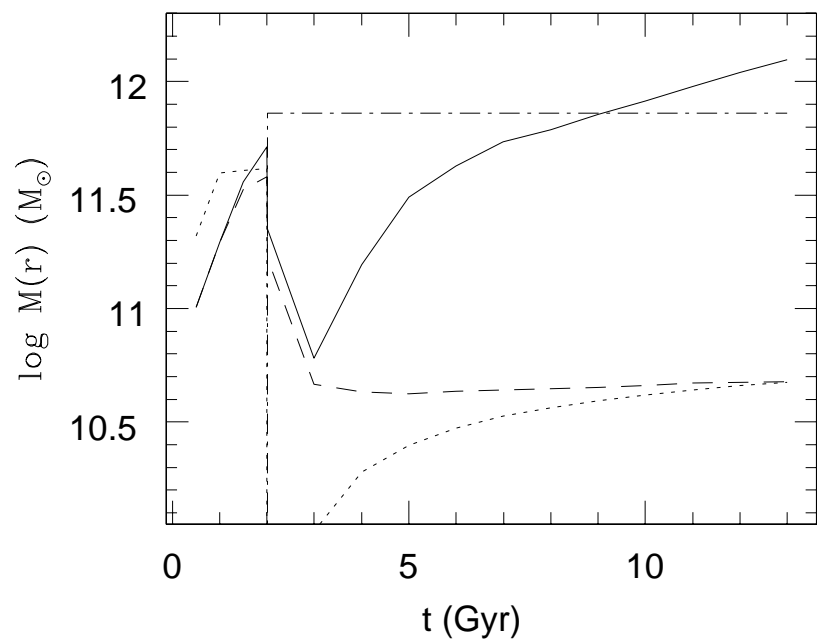
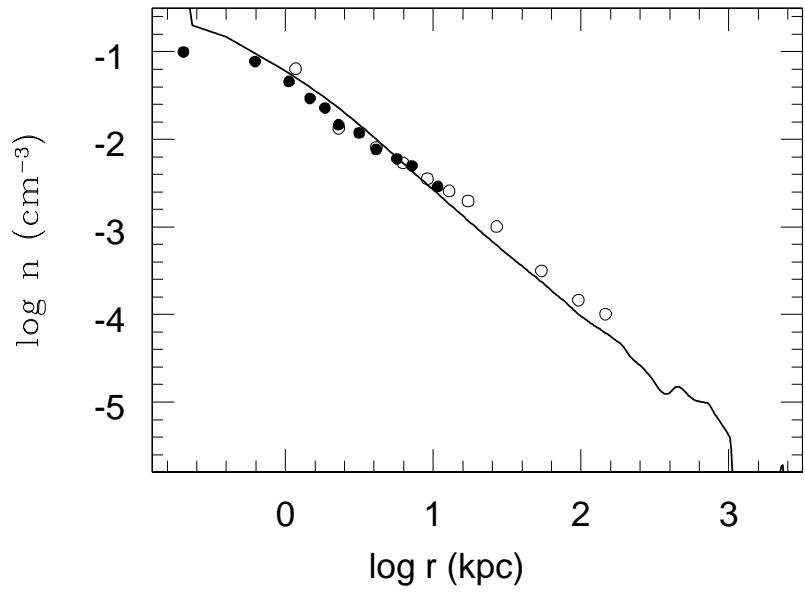
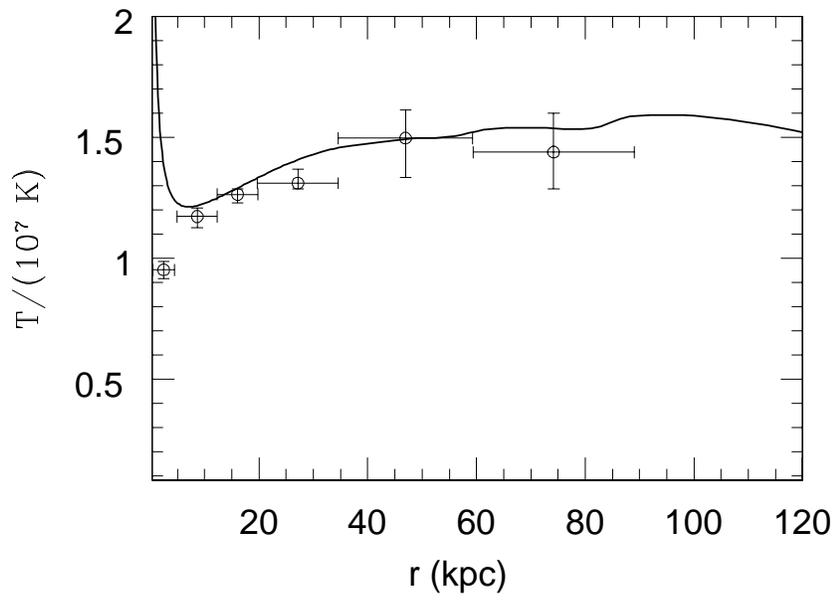
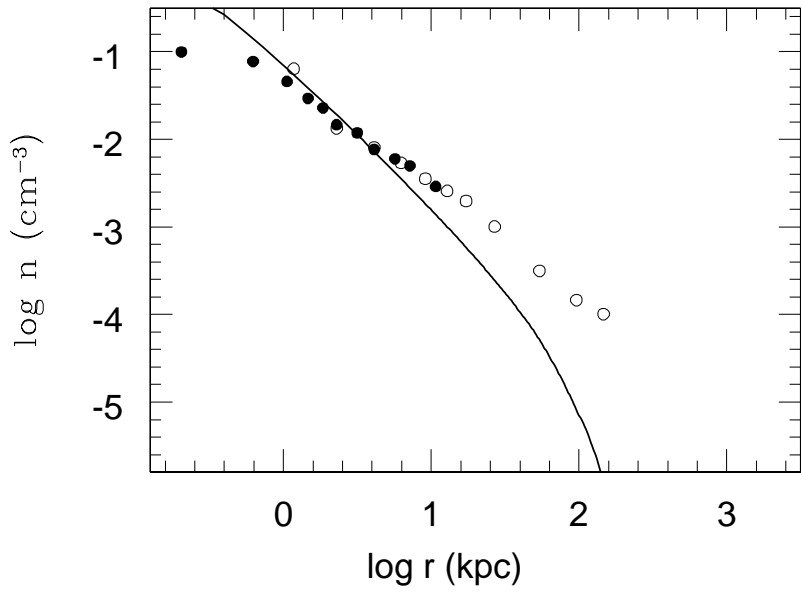
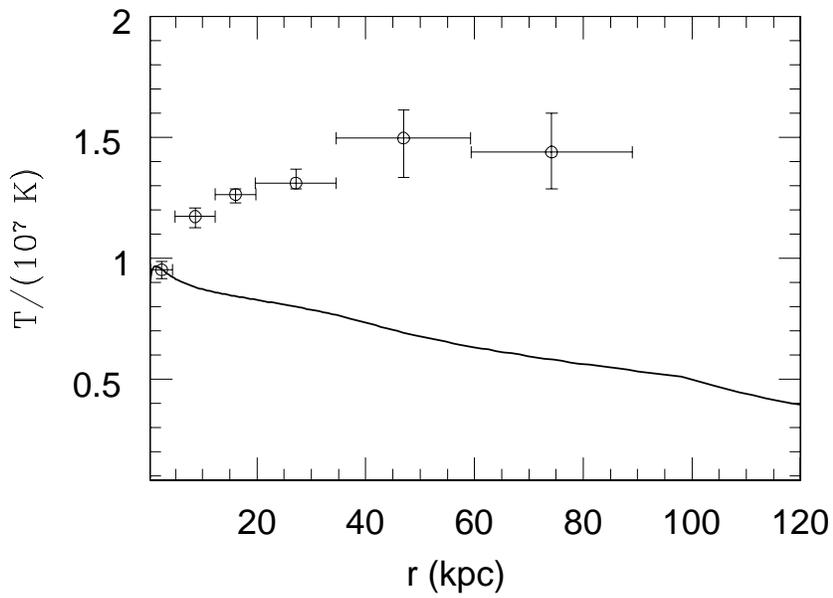
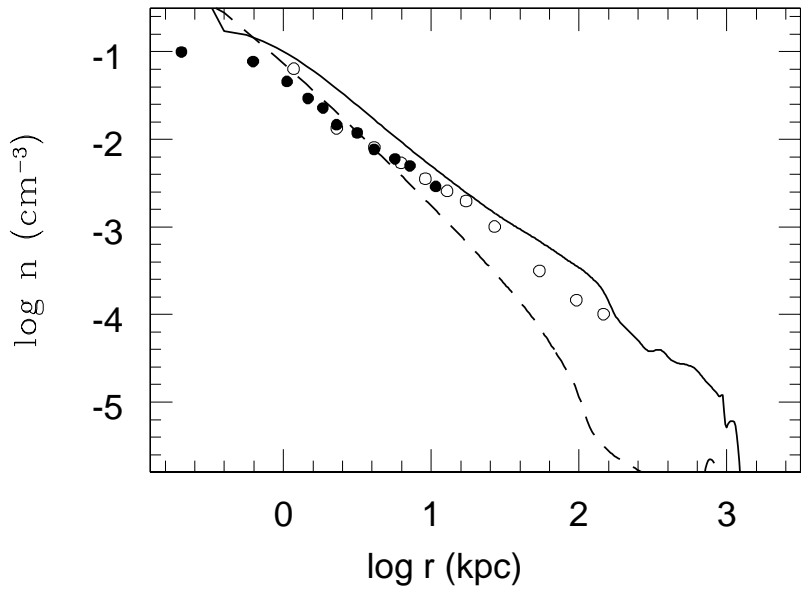
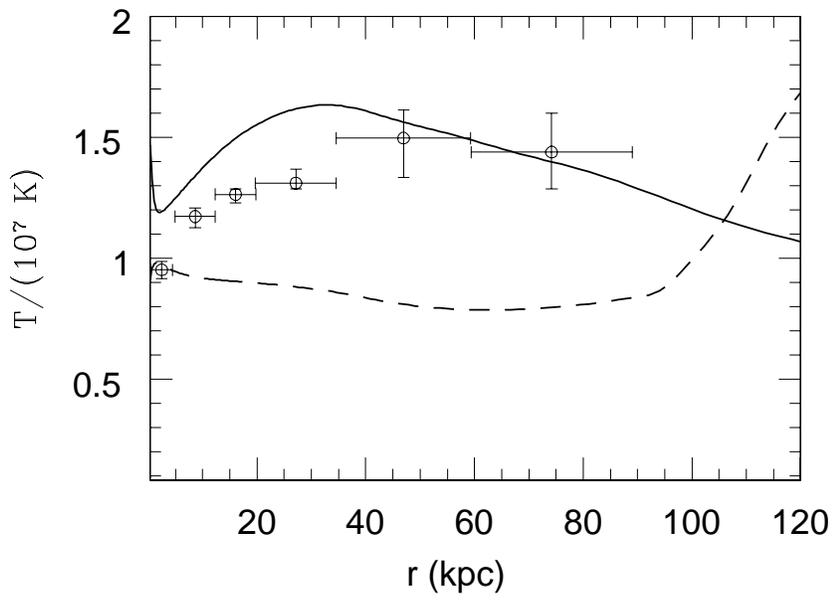
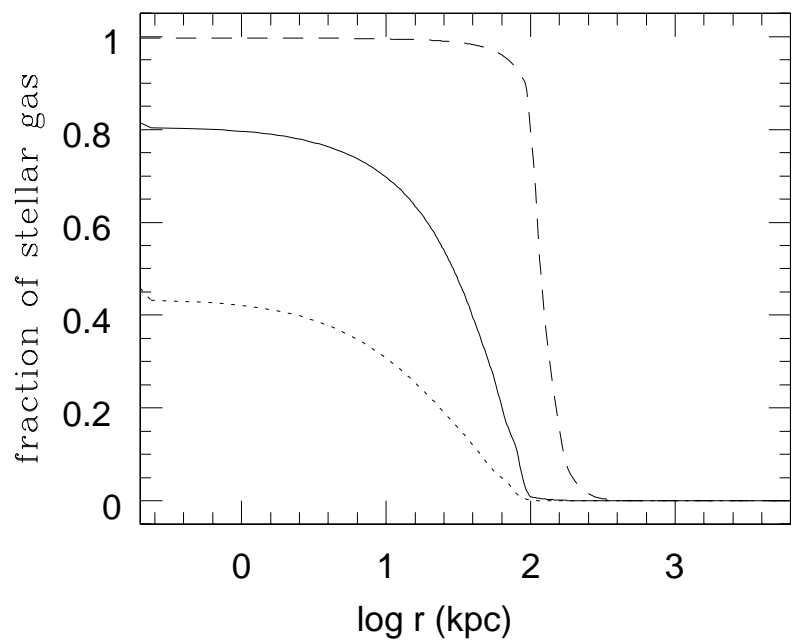
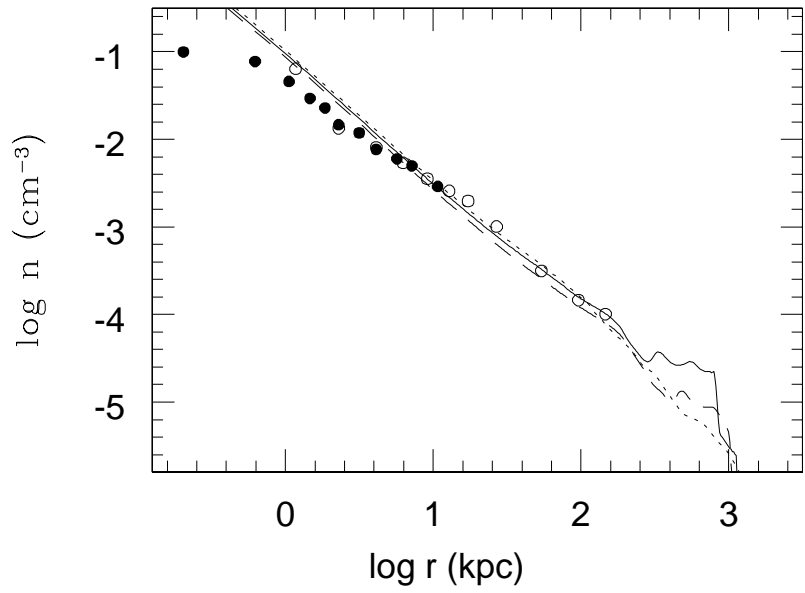
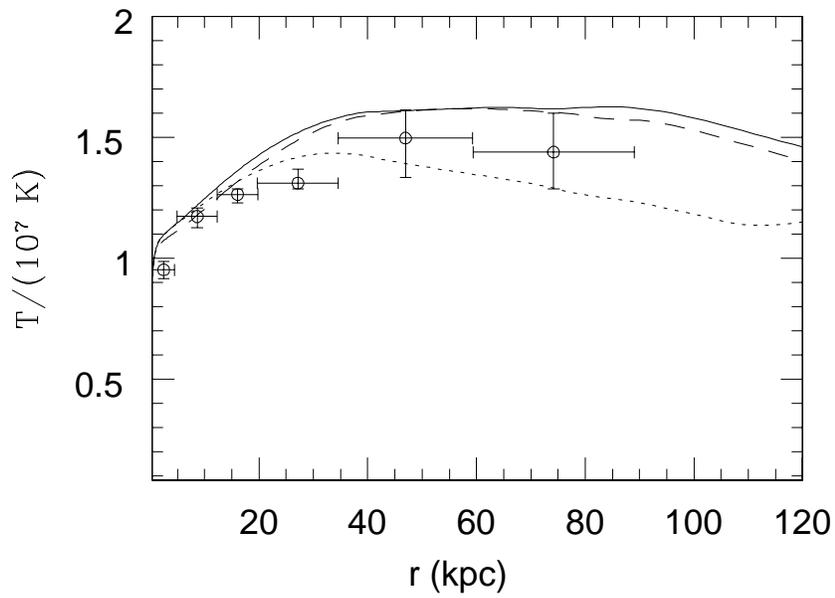
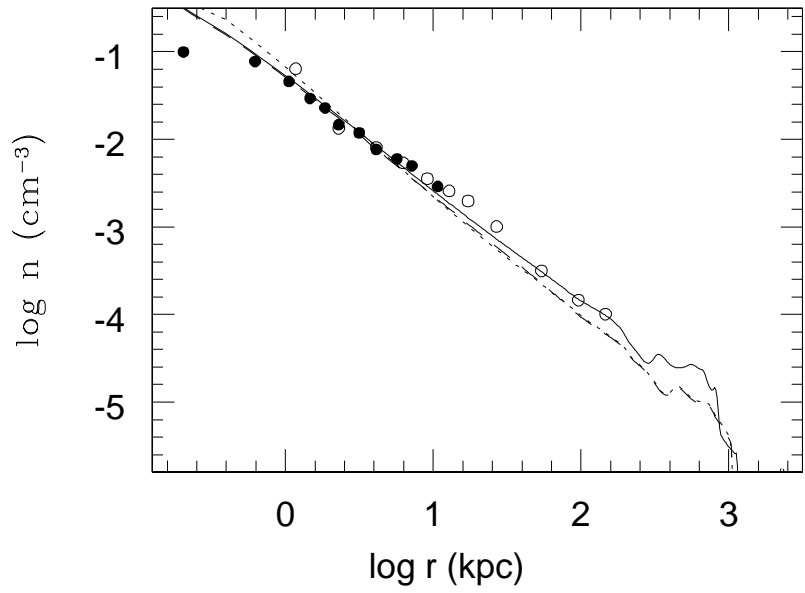
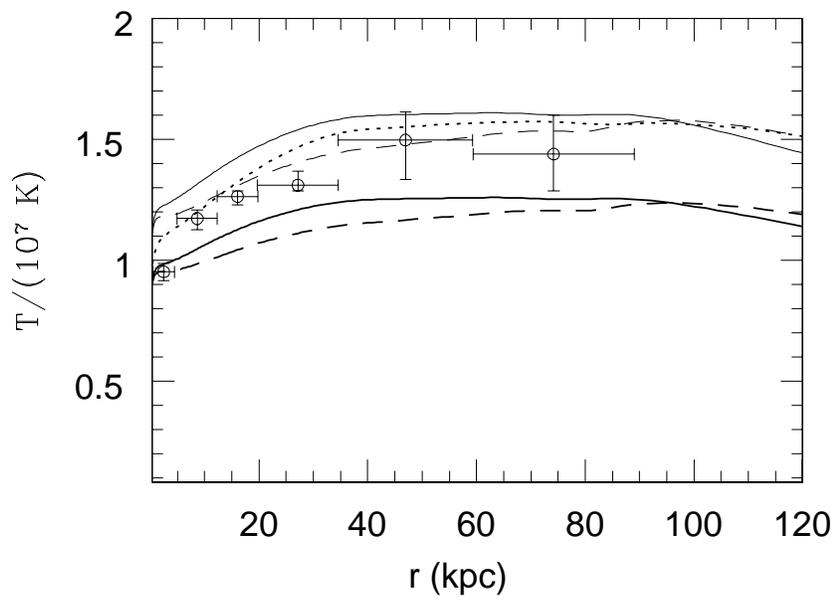
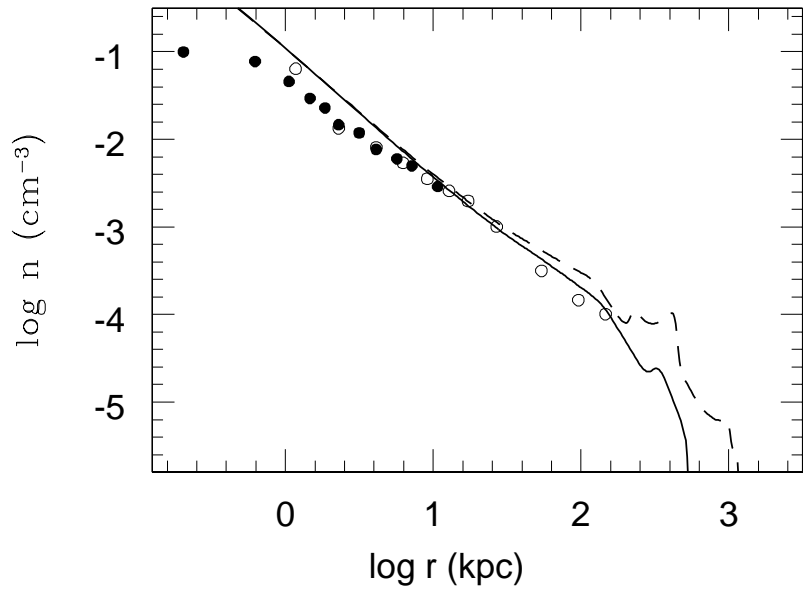
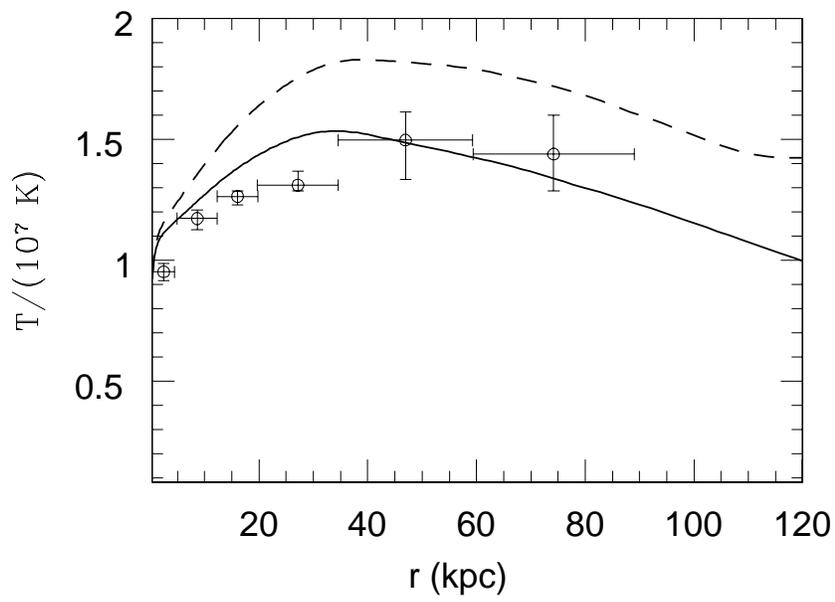
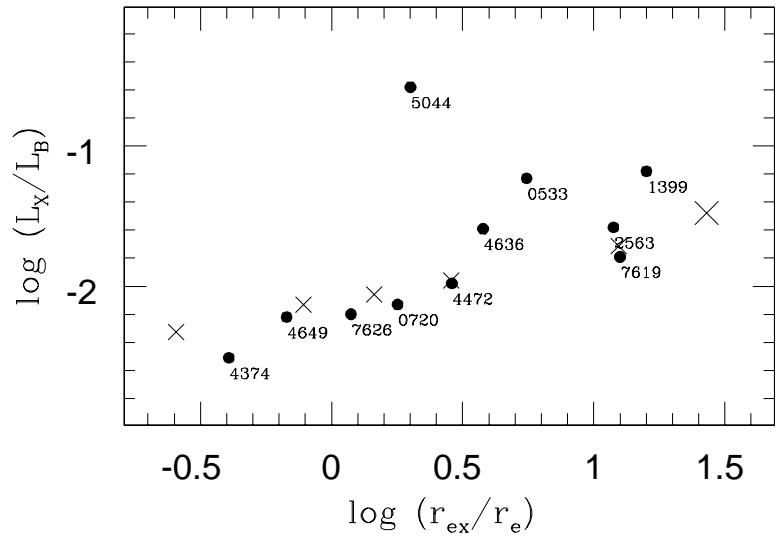
\end{document}